\PassOptionsToPackage{dvipsnames}{xcolor}
\documentclass[preprints,article,accept,pdftex,moreauthors]{Definitions/mdpi} 

\firstpage{1} 
\pubvolume{1}
\issuenum{1}
\articlenumber{0}
\pubyear{2024}
\copyrightyear{2024}
\datereceived{ } 
\daterevised{ } 
\dateaccepted{ } 
\datepublished{ } 
\hreflink{https://doi.org/} 

\usepackage{xspace}
\usepackage{siunitx}
\usepackage{subcaption}
\usepackage{tikz}
\usetikzlibrary{arrows, angles, quotes, decorations.pathreplacing, calligraphy, calc}
\usepackage{import}

\DeclareRobustCommand{\Pnu}{$\nu$\xspace}
\DeclareRobustCommand{\Pnue}{$\nu_\mathrm{e}$\xspace}
\DeclareRobustCommand{\Pnum}{$\nu_\mu$\xspace}

\DeclareRobustCommand{\chips}{CHIPS\xspace}
\DeclareRobustCommand{\chipsfive}{CHIPS-5\xspace}
\DeclareRobustCommand{\nemo3}{NEMO-3\xspace}
\DeclareRobustCommand{\km3net}{KM3NeT\xspace}
\DeclareRobustCommand{\numi}{NuMI\xspace}
\DeclareRobustCommand{\daq}{DAQ\xspace}
\DeclareRobustCommand{\promis}{PROMiS\xspace}
\DeclareRobustCommand{\coco}{CoCo\xspace}
\DeclareRobustCommand{\isquaredc}{I\textsuperscript{2}C\xspace}
\DeclareRobustCommand{\cat5}{Cat~5\xspace}
\DeclareRobustCommand{\microdaq}{MicroDAQ\xspace}
\DeclareRobustCommand{\bb}{BeagleBone\textsuperscript{\tiny\textregistered}\xspace}
\DeclareRobustCommand{\bbg}{\bb~Green\xspace}

\DeclareSIUnit{\inch}{\text{in}}

\foreach \x in {A, ..., Z}{%
\expandafter\xdef\csname orcidA\x\endcsname{\noexpand\href{https://orcid.org/\csname orcidauthorA\x\endcsname}{\noexpand\orcidicon}}
}

\Title{Scalable DAQ system operating the CHIPS-5 neutrino detector}

\TitleCitation{Scalable DAQ system operating the CHIPS-5 neutrino detector}

\Author{%
Belén~Alonso~Rancurel$^{2}$,
Nicolas~Angelides$^{5}$,
Guillaume~Augustoni$^{2}$,
Simeon~Bash$^{1,\dagger}$\orcidB{},
Benedikt~Bergmann$^{3}$\orcidC{},
Nicholas~Bertschinger$^{4}$,
Paul~Bizouard$^{1}$,
Medbh~Campbell$^{1}$,
Son~Cao$^{19}$\orcidY{},
Thomas~J.~Carroll$^{4}$,
Rhys~Castellan$^{1}$,
Erika~Catano-Mur$^{14}$\orcidD{},
John~P.~Cesar$^{12}$\orcidE{},
João~A.~B.~Coelho$^{22}$\orcidF{},
Patrick~Dills$^{28}$,
Thomas~Dodwell$^{1}$ ,
Jack~Edmondson$^{1}$,
Daan~van~Eijk$^{7}$,
Quinn~Fetterly$^{4}$,
Zoé~Garbal$^{2}$,
Stefano~Germani$^{1,23}$,
Thomas~Gilpin$^{1}$,
Anthony~Giraudo$^{2}$,
Alec~Habig$^{11}$\orcidG{},
Daniel~Hanuska$^{4}$,
Harry~Hausner$^{9}$\orcidH{},
Wilson~Y.~Hernandez$^{4}$,
Anna~Holin$^{20}$\orcidI{},
Junting~Huang$^{18}$,
Sebastian~B.~Jones$^{1}$\orcidJ{},
Albrecht~Karle$^{16}$\orcidK{},
George~Kileff$^{1}$,
Kai~R.~Jenkins$^{1}$,
Paul~Kooijman$^{7}$,
Arthur~Kreymer$^{9}$\orcidL{},
Gabe~M.~LaFond$^{4}$,
Karol~Lang$^{12}$\orcidM{},
Jeffrey~P.~Lazar$^{16}$\orcidN{},
Rui~Li$^{10}$,
Kexin~Liu$^{24,25}$,
David~A.~Loving$^{14}$,
Petr~Mánek*$^{,1,3}$\orcidA{},
Marvin~L.~Marshak$^{13}$\orcidO{},
Jerry~R.~Meier$^{13}$,
William~Miller$^{13}$,
Jeffrey~K. Nelson$^{14}$\orcidP{},
Christopher~Ng$^{27}$,
Ryan~J.~Nichol$^{1}$\orcidQ{},
Vittorio~Paolone$^{21}$\orcidR{},
Andrew~Perch$^{1}$,
Maciej~M.~Pfützner$^{1}$,
Alexander~Radovic$^{1,14}$,
Katherine~Rawlins$^{8}$,
Patrick~Roedl$^{4}$,
Lucas~Rogers$^{4}$,
Ibrahim~Safa$^{15}$\orcidS{},
Alexandre~Sousa$^{17}$\orcidT{},
Josh~Tingey$^{1}$\orcidZ{},
Jennifer~Thomas$^{1}$\orcidAA{},
Jozef~Trokan-Tenorio$^{14}$,
Patricia~Vahle$^{14}$\orcidU{},
Richard~Wade$^{6}$,
Christopher~Wendt$^{4}$\orcidV{},
Daniel~Wendt$^{26}$,
Leigh~H.~Whitehead$^{1,\ddagger}$\orcidW{},
Samuel~Wolcott$^{4}$,
Tianlu~Yuan$^{16}$\orcidX{}%
}

\AuthorNames{Rancurel Belén Alonso, Angelides Nicolas, Augustoni Guillaume, Bash Simeon, Bergmann Benedikt, Bertschinger Nicholas, Bizouard Paul, Campbell Medbh, Cao Son, Carroll Thomas J., Castellan Rhys, Catano-Mur Erika, Cesar John P., Coelho João A. B., Dills Patrick, Dodwell Thomas, Edmondson Jack, van Eijk Daan, Fetterly Quinn, Garbal Zoé, Germani Stefano, Gilpin Thomas, Giraudo Anthony, Habig Alec, Hanuska Daniel, Hausner Harry, Hernandez Wilson Y., Holin Anna, Huang Junting, Jones Sebastian B., Karle Albrecht, Kileff George, Jenkins Kai R., Kooijman Paul, Kreymer Arthur, LaFond Gabe M., Lang Karol, Lazar Jeffrey P., Li Rui, Liu Kexin, Loving David A., Mánek Petr, Marshak Marvin L., Meier Jerry R., Miller William, Nelson Jeffrey K., Ng Christopher, Nichol Ryan J., Paolone Vittorio, Perch Andrew, Pfützner Maciej M., Radovic Alexander, Rawlins Katherine, Roedl Patrick, Rogers Lucas, Safa Ibrahim, Sousa Alexandre, Tingey Josh, Thomas Jennifer, Trokan-Tenorio Jozef, Vahle Patricia, Wade Richard and Yuan Tianlu}

\AuthorCitation{Mánek P. et al}

\address{%
$^{1}$ \quad Department of Physics and Astronomy, University College London, Gower Street, London, WC1E 6BT, United Kingdom\\
$^{2}$ \quad Aix-Marseille University, Science Faculty in Saint-Jérôme campus, 13013 Marseille, France\\
$^{3}$ \quad Institute of Experimental and Applied Physics, Czech Technical University in Prague, Husova 240/5, 110 00  Prague, Czech Republic\\
$^{4}$ \quad Department of Physics, University of Wisconsin, Madison, WI 53706, United States\\
$^{5}$ \quad Imperial College London, Physics Department, Blackett Laboratory, London SW7 2AZ, United Kingdom\\
$^{6}$ \quad Avenir Consulting, Abingdon, Oxfordshire, United Kingdom\\
$^{7}$ \quad Nikhef, Science Park 105, 1098 XG, Amsterdam, Netherlands\\
$^{8}$ \quad University of Alaska Anchorage, 3211 Providence Dr. Anchorage, AK 99508, United States\\
$^{9}$ \quad Fermi National Accelerator Laboratory, Batavia, IL 60510, United States\\
$^{10}$ \quad School of Physics and Astronomy, Shanghai Jiao Tong University, MOE Key Laboratory for Particle Astrophysics and Cosmology, Shanghai Key Laboratory for Particle Physics and Cosmology, Shanghai 200240, China\\
$^{11}$ \quad Department of Physics and Astronomy, University of Minnesota Duluth, Duluth, MN 55812, United States\\
$^{12}$ \quad Department of Physics, University of Texas at Austin, Austin, TX 78712, United States\\
$^{13}$ \quad University of Minnesota, Minneapolis, MN 55455, United States\\
$^{14}$ \quad Department of Physics, William \& Mary, Williamsburg, VA 23187, United States\\
$^{15}$ \quad Department of Physics, Columbia University, New York, NY, United States\\
$^{16}$ \quad Department of Physics and Wisconsin IceCube Particle Astrophysics Center, University of Wisconsin–Madison, Madison, WI 53706, United States\\
$^{17}$ \quad Department of Physics, University of Cincinnati, Cincinnati, OH 45221, United States\\
$^{18}$ \quad School of Physics and Astronomy, Shanghai Jiao Tong University, China\\
$^{19}$ \quad Institute For Interdisciplinary Research in Science and Education, ICISE, Quy Nhon, Vietnam\\
$^{20}$ \quad Particle Physics Department, STFC Rutherford Appleton Laboratory, Harwell Campus,  Didcot OX11 0QX, United Kingdom\\
$^{21}$ \quad University of Pittsburgh, Pittsburgh, PA 15260, United States\\
$^{22}$ \quad Université Paris Cité, Astroparticule et Cosmologie, F-75013 Paris, France\\
$^{23}$ \quad Dipartimento di Fisica e Geologia, Università degli Studi di Perugia, I-06123 Perugia, Italy\\
$^{24}$ \quad Department of Physics, Sun Yat-sen University, 135 Xingang Xi Road, Haizhu District, Guangzhou 510275, China\\
$^{25}$ \quad Department of Physics, Tsinghua University, 30 Shuangqing Road, Haidian District, Beijing 100084, China\\
$^{26}$ \quad Department of Physics, University of Colorado, Boulder, CO 80309, United States\\
$^{27}$ \quad Department of Physics and Astronomy, Michigan State University, East Lansing, MI 48824, United States\\
$^{28}$ \quad Department of Mechanical Engineering, University of Wisconsin, Madison, WI 53706, United States%
}

\corres{Correspondence: petr.manek.19@ucl.ac.uk}

\firstnote{Now at Technische Universität München, James-Franck-Straße, 85748 Garching, Germany}
\secondnote{Now at Cavendish Laboratory, University of Cambridge, Cambridge CB3 0HE, United Kingdom}

\abstract{The CHIPS R\&D project focuses on development of low-cost water Cherenkov neutrino detectors through novel design strategies and resourceful engineering. This work presents an end-to-end DAQ~solution intended for a recent 5~kt CHIPS prototype, which is largely based on affordable mass-produced components. Much like the detector itself, the presented instrumentation is composed of modular arrays that can be scaled up and easily serviced. A single such array can carry up to 30~photomultiplier tubes~(PMTs) accompanied by electronics that generate high voltage in-situ and deliver time resolution of up to 0.69~ns. In addition, the technology is compatible with the White Rabbit timing system, which can synchronize its elements to within 100~ps. While deployment issues did not permit the presented DAQ~system to operate beyond initial evaluation, the presented hardware and software successfully passed numerous commissioning tests that demonstrated their viability for use in a large-scale neutrino detector, instrumented with thousands of PMTs.}


\begin{document}

\section{Introduction} 
\label{sec:intro} 

Experiments studying neutrinos address low signal rates by constructing massive detectors that can collect statistically significant data within acceptable time frame. Building a large apparatus on the scale of tens or hundreds of kilotons, however, poses a series of unique challenges that call for large research collaborations and prohibitively substantial upfront investments. The Cherenkov Detectors in Mine Pits (\chips) R\&D~project~\cite{adamson2013letterofintent} sidesteps many such issues by developing low-cost neutrino detector technology based on affordable mass-produced components. \chips-style detectors employ modular architecture, which allows them to be readily constructed at small scale and expanded on demand, easily serviced, and operated at multiple locations to observe coincident events with sub-nanosecond time resolution.

\begin{figure}
    \centering
    \input{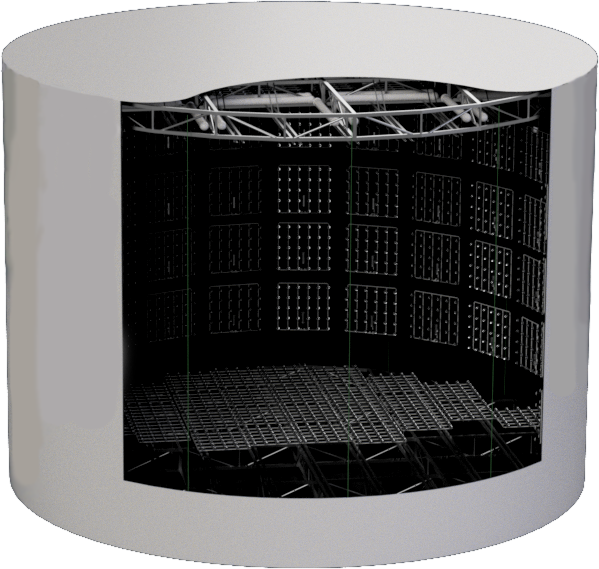}
    \caption{Rendering of \chipsfive in its fully deployed state, courtesy of T.~Dodwell.}
    \label{fig:chips:overview:parts}
\end{figure}

\begin{figure}
    \centering
    {
        \def\svgwidth{\columnwidth}
        \small
        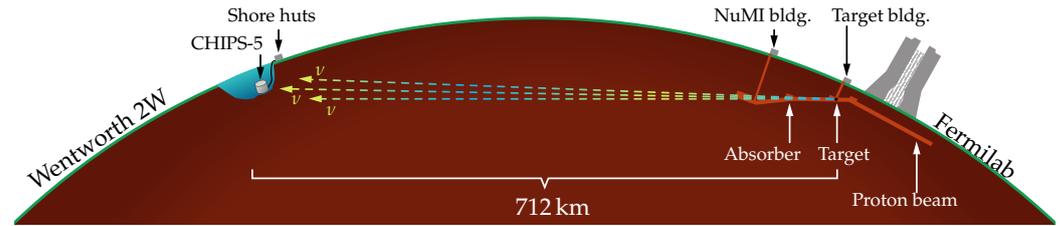
    }
    \caption{Intended operation of \chipsfive in the path of the \numi neutrino beam (not to scale). Neutrinos generated at Fermilab (right) travel through Earth's crust to emerge at the Wentworth 2W mining pit, where \chipsfive was deployed (left).}
    \label{fig:chips:overview:schematic}
\end{figure}

To demonstrate and verify this concept, \chips tested a number of detector prototypes over the last decade. The latest and largest one to date was \chipsfive, which was realized between~2018 and~2020~\cite{rancurel2024design}. \chipsfive (pictured in \Cref{fig:chips:overview:parts}) was a \qty[round-precision=1]{5.9}{\kilo\tonne} water Cherenkov detector that comprised a cylindrical tank, \qty[round-precision=0]{12}{\meter}~tall and \qty[round-precision=0]{25}{\meter}~across, which was deployed underwater in a disused mining pit near Hoyt Lakes, Minnesota, \qty[round-precision=0]{7}{\milli\radian}~off axis of the \numi~neutrino beam~\cite{adamson2016numi}. This location allowed \chipsfive to study neutrino oscillations by probing \Pnum~disappearance over a \qty[round-precision=0]{712}{\kilo\meter}~baseline, as illustrated in \Cref{fig:chips:overview:schematic}.

Modular aspects of \chipsfive's architecture were by no means limited only to its structural elements. The detector in its entirety was designed to be expandable, so that its technology may be used in larger-scale deployments without significant alteration. To this end, its photosensitive instrumentation was organized into planar optical modules~(POMs) -- segmented arrays of photomultiplier tubes~(PMTs) that were mounted on its internal surface. Each POM was an autonomous measuring unit that in addition to PMTs carried electronics for high voltage generation, time synchronization and pulse digitization. Interlinked POMs gave rise to a network that was controlled by dedicated software, which aggregated data and performed clustering in real-time. Together, these components formed a data acquisition~(\daq) system, which facilitated flow of data from PMTs to a persistent storage facility at Fermilab.

While it is infeasible to give a full account of the detector design details in the scope of a single article, this work aims to provide an overview of its functional components with principal focus on the \daq system. The rest of the text is therefore structured as follows. \Cref{sec:daq:hardware} is devoted to low-level electronics, such as PMTs and POMs, which were responsible for digitizing and timestamping measurements. Next, \Cref{sec:daq:infra-hardware} discusses infrastructure that distributed power, data links and precise timing signals to all elements of the system. \Cref{sec:daq:software} is devoted to the software component of the \daq system, which serviced the hardware, interacted with operators and performed data analysis. Finally, \Cref{sec:conclusion} summarizes outcomes of the \chipsfive deployment and future prospects of its \daq technology.

\section{Measurement hardware} 
\label{sec:daq:hardware} 

Even though \chipsfive's instrumentation shared the same basic structure and purpose, its hardware was not homogeneous. POMs were divided into two functionally equivalent groups, labeled by their institutes of origin: Nikhef and University of Wisconsin, Madison. While Nikhef POMs incorporated electronics that were already in use at the \km3net experiment, Madison POMs were seen as a proving ground for a modern PMT readout system developed by \chips in collaboration with Wisconsin IceCube Particle Astrophysics Center~(WIPAC). This permitted the newer Madison electronics to be qualified by comparison with a well-characterized Nikhef system. Under full instrumentation, \chipsfive was designed to hold \num[round-precision=0]{226}~Nikhef POMs and \num[round-precision=0]{30}~Madison POMs.



Irrespective of their underlying technology, POMs observed photons through PMTs, which produced analog pulses corresponding to registered light. POM hardware was therefore mainly responsible for supplying PMTs with high voltage, which determined their gain, and digitizing their pulses. Since thousands of PMTs were planned to be instrumented, hardware servicing a single channel needed to remain relatively small and inexpensive. This motivated a tree-shaped architecture, where the lowest-level elements were plain, mass-produced circuit boards. At higher levels of the tree, components became more sophisticated and less numerous, as they aggregated data from lower stages. In addition, timing played a critical role. To be able to correlate hits between multiple channels, digitization mechanism required precise synchronization throughout the entire system. Considering detector dimensions, light was expected to traverse its entire volume within approximately~\qty[round-precision=1]{83.39102379953801}{\nano\second}. This set the time synchronization target to be on the order of nanoseconds, so that sources of Cherenkov light may be accurately localized.

\subsection{Photomultiplier tubes} 
\label{sec:daq:pmts} 

At the lowest level of the \daq~system, \chipsfive used two distinct PMT~models that aligned with POM types. Their relevant parameters are listed in \Cref{tab:daq:pmts:parameters}. Nikhef POMs used the XP82B20FNB~model manufactured by HZC, a Chinese company that recently acquired the photomultiplier division of Photonis. These PMTs had a domed photocathode and measured approximately~\qty[round-precision=0]{88}{\milli\meter} in diameter~\cite{hzc2019datasheet}. In contrast, Madison POMs carried Hamamatsu~R6091 PMTs, which were donated from the \nemo3 experiment~\cite{arnold2005technical}. Their photocathode was flat and approximately~\qty[round-precision=0]{76}{\milli\meter} across~\cite{hamamatsu2016datasheet}. Both PMT~models were selected to offer acceptable response around~\qty[round-precision=0]{400}{\nano\meter}, which was the estimated wavelength of expected Cherenkov light from neutrinos.

\begin{table}
    \begin{tabular*}{\columnwidth}{@{\extracolsep{\fill}} l|rr}
    \toprule
    Parameter & HZC XP82B20FNB~\cite{hzc2019datasheet} & Hamamatsu R6091~\cite{hamamatsu2016datasheet}\\
    \midrule
    Diameter, length & \qty[round-precision=0]{88}{\milli\meter}, \qty[round-precision=0]{94.5}{\milli\meter} & \qty[round-precision=0]{76}{\milli\meter}, \qty[round-precision=0]{137}{\milli\meter}\\
    Gain & \num[round-precision=0]{1e7} & \num[round-precision=0]{5e6}\\
    Anode pulse rise time & \qty[round-precision=1]{3.5}{\nano\second} & \qty[round-precision=1]{2.6}{\nano\second}\\
    Quantum efficiency at \qty[round-precision=0]{400}{\nano\meter} & \qty[round-precision=1]{24.4}{\percent} & \qty[round-precision=1]{12.0}{\percent}\\
    Typical (max) anode dark current & \qty[round-precision=0]{20}{\nano\ampere} (\qty[round-precision=0]{300}{\nano\ampere}) & \qty[round-precision=0]{10}{\nano\ampere} (\qty[round-precision=0]{60}{\nano\ampere}) \\
    \bottomrule
    \end{tabular*}
    \caption{Selected parameters of PMTs, which were used in \chipsfive.}
    \label{tab:daq:pmts:parameters}
\end{table}

Under normal operation PMTs required steady supply of high voltage in the range of \qtyrange[round-precision=1]{1}{2.5}{\kilo\volt}. Since this potential was proportional to gain, it was desirable to adjust it for each channel independently in order to equalize response across the entire detector. In addition, it was viewed as hazardous to distribute high voltages underwater from a centralized supply. PMTs were therefore fitted with bases that generated high voltage in-situ using a Cockroft-Walton~(CW) voltage multiplier~\cite{cockcroft1930experiments}. This way, power lines could supply~PMTs with much safer low-voltage direct current, which was switched and modulated by electronics into high voltages just prior to consumption.

Even though both PMT models shared similar characteristics, their installation in \chipsfive required distinct treatment due to differences in polarity. Hamamatsu~PMTs were wired with anodes at high voltage and photocathodes close to ground, which permitted the photocathodes to be directly exposed to pit water without any undesirable effects. In contrast, HZC~PMTs used inverted configuration with photocathodes at high voltage, which would introduce additional noise into measurements if left unaddressed. For that reason, HZC~photocathodes were electrically insulated from surrounding water by transparent acrylic shields that were installed during potting. In the course of this procedure, a gelatinous compound was injected into \qty[round-precision=1]{0.5}{\milli\meter}~gap between a~PMT and its shield, and left to harden. To minimize light losses due to reflection and refraction at the optical interface, the compound was selected to match refractive indices of adjacent materials, and the procedure was performed under vacuum that removed any bubbles. A dedicated workflow was devised to efficiently pot batches of PMTs in groups of~155.

\begin{figure}
    \centering
    \begin{subfigure}{0.48\columnwidth}
        \includegraphics[width=\textwidth]{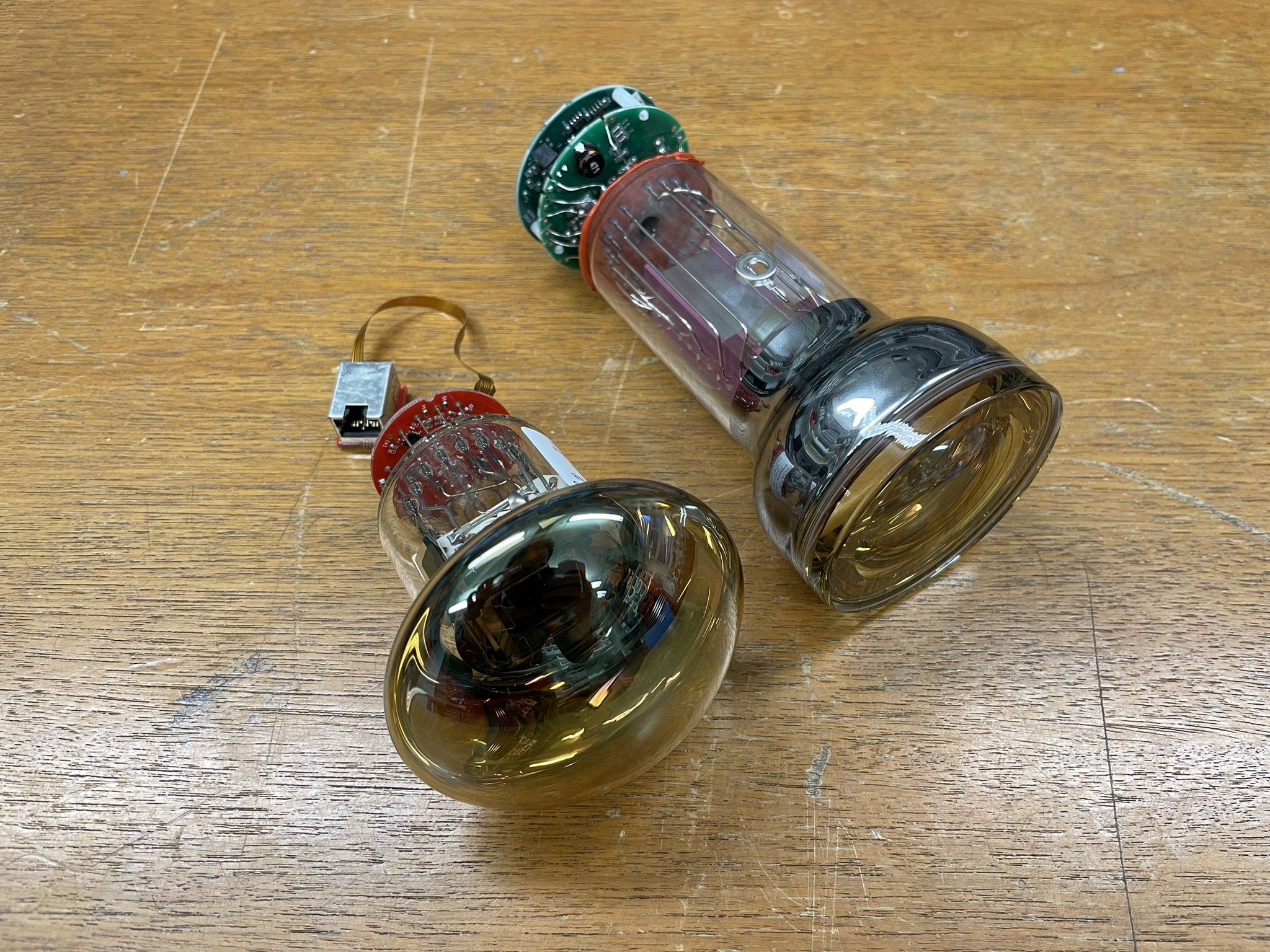}
        \caption{PMTs with bases before waterproofing}
        \label{fig:daq:pmts:assembly:before}
    \end{subfigure}%
    \quad%
    \begin{subfigure}{0.48\columnwidth}
        \includegraphics[width=\textwidth]{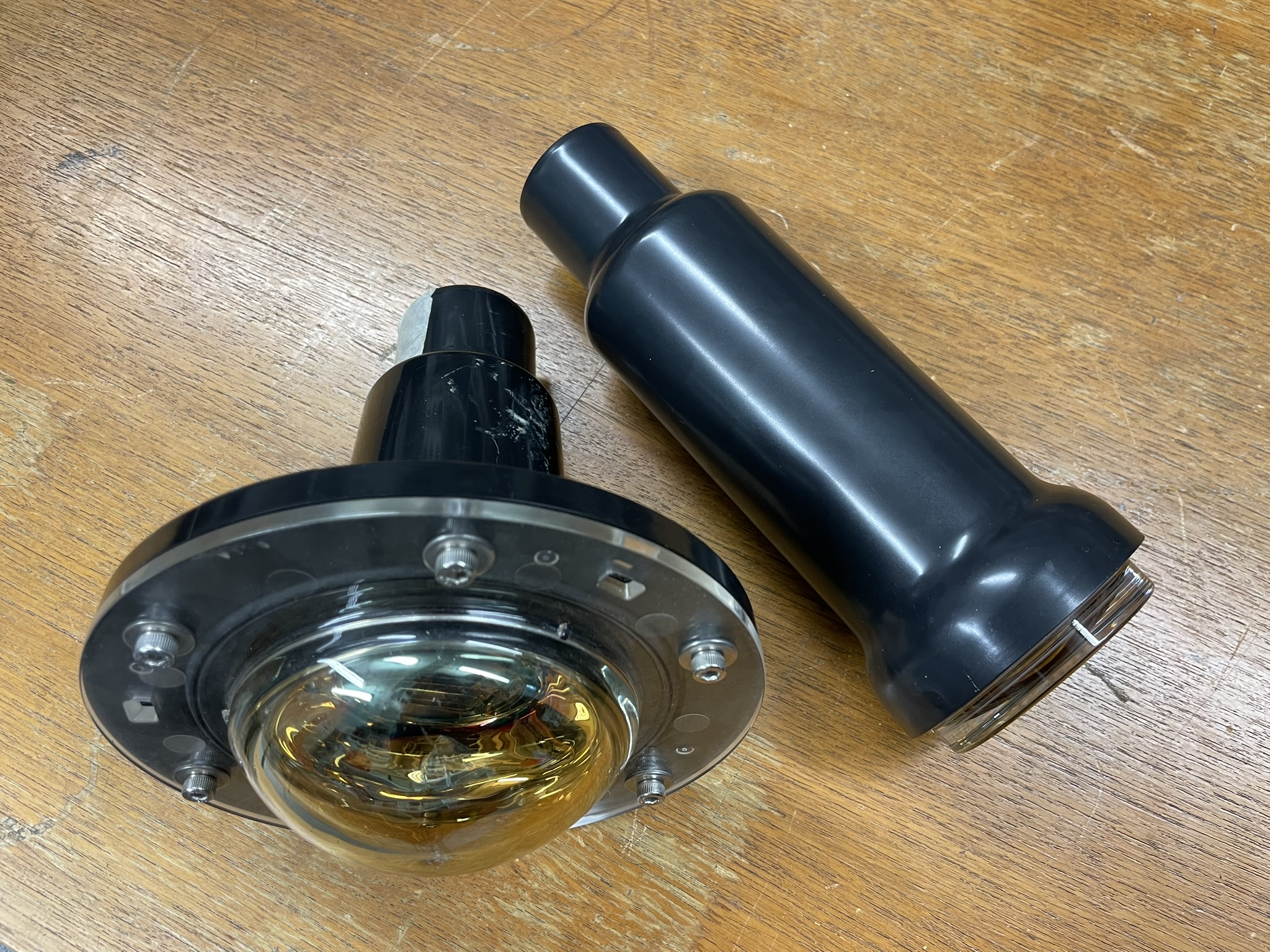}
        \caption{PMTs inside cases, ready for installation}
        \label{fig:daq:pmts:assembly:after}
    \end{subfigure}
    \caption{Waterproofing of PMTs before their installation in POMs. Photographs show PMTs manufactured by HZC~(left) and Hamamatsu~(right) before and after the procedure. Note the domed acrylic shield on the HZC~PMT that insulates its photocathode.}
    \label{fig:daq:pmts:assembly}
\end{figure}

In final stages of installation PMTs were mechanically affixed to POMs by plastic inserts that formed water-tight seals with their PVC~tubing. To prevent water ingress in photocathode area, PMTs underwent different procedures based on their model. HZC~PMT cases were designed to completely enclose their PMTs, so inserts covered the entire tube and interfaced with its acrylic shield by means of o-rings and screws. In contrast, Hamamatsu~PMT cases were designed to leave photocathodes exposed, so they were permanently attached to the rim of the tube by water-resistant sealant. \Cref{fig:daq:pmts:assembly} shows completed assemblies for both PMT models.

\subsection{Nikhef POMs} 
\label{sec:daq:nikhef_poms} 

POMs of the Nikhef type were the most abundant in the \daq~system. In the first phase of \chipsfive deployment, both detector endcaps were instrumented with \num[round-precision=0]{226}~Nikhef POMs carrying \num[round-precision=0]{6114}~PMTs. Of this number, \num[round-precision=0]{324}~PMTs located in the top endcap were oriented upward to observe the veto region and assist with cosmic background rejection. The rest were oriented into the inner volume.

Each Nikhef POM measured approximately~\qtyproduct[round-precision=0]{2 x 3}{\meter} and carried up to 30~HZC~PMTs, which were driven by electronics developed for the \km3net project~\cite{katz2009status,adrian2016letter}. These components can be viewed as two specific groups: application-specific integrated circuits~(ASICs) that were embedded in PMT~bases, and printed circuit boards~(PCBs) that were installed in a cylindrical container in the middle of the POM. Under normal operation data flowed from~ASICs to~PCBs, which aggregated measurements from the entire POM and transmitted them upstream to higher stages of the \daq~system.

\begin{figure}
    \centering
    \input{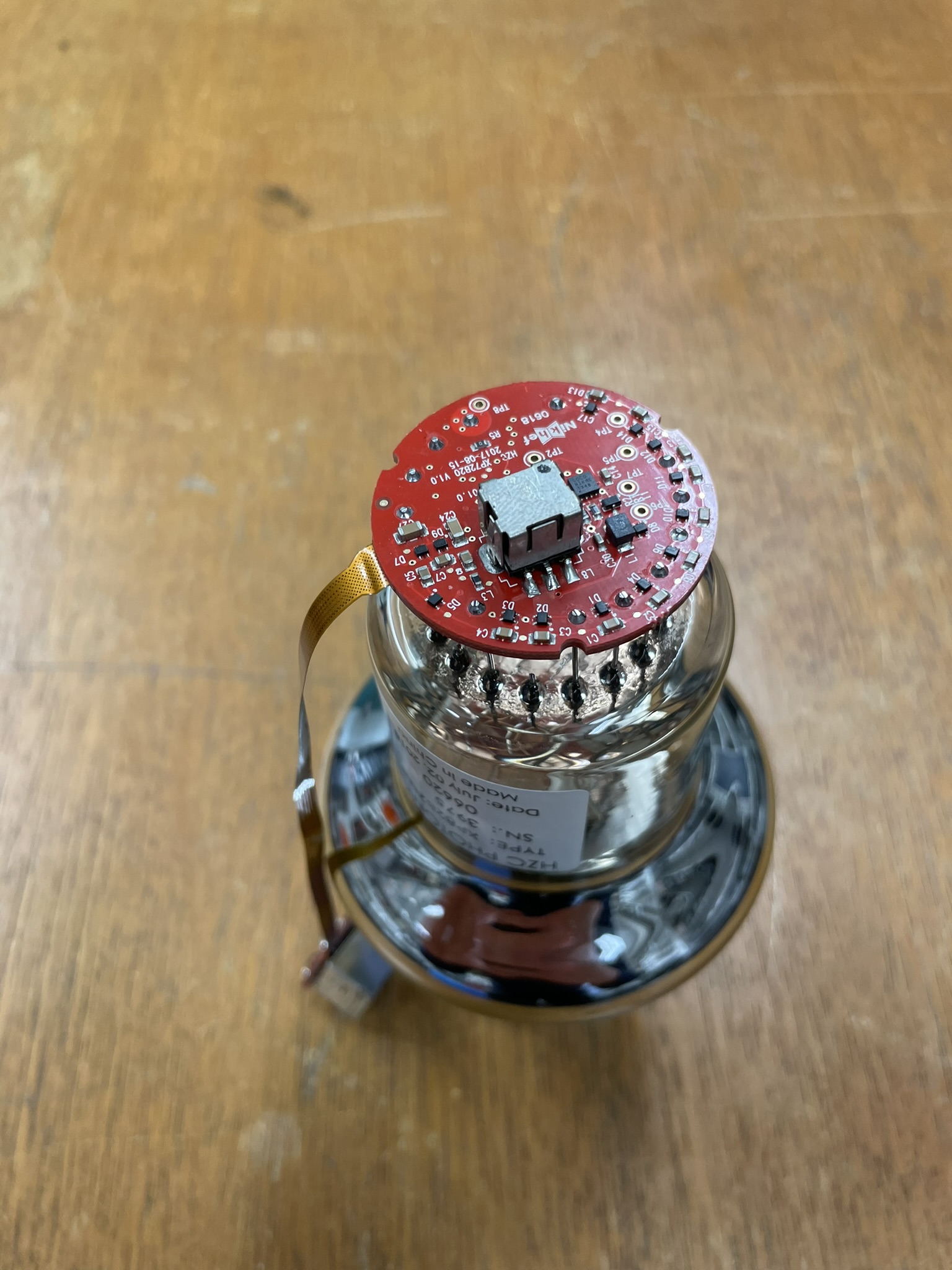}
    \caption{High voltage base of a HZC~PMT carrying~ASICs. While \coco is visible in the photograph, \promis is fitted on the underside, out of view.}
    \label{fig:daq:nikhef_poms:coco}
\end{figure}

A base of an HZC~PMT is shown in \Cref{fig:daq:nikhef_poms:coco}. Its high voltage was generated by an~ASIC known as \coco~(Cockroft-Walton Multiplier Feedback Control)~\cite{gajanana2013asic}. This circuit produced periodic pulses of frequencies up to~\qty[round-precision=0]{50}{\kilo\hertz} that drove a CW~voltage multiplier. As its name suggests, \coco included a sensing feedback mechanism that allowed it to adaptively regulate voltage, so that stable PMT~gain could be maintained at desired levels over extended periods of time.

\begin{figure}
    \centering
    {
        \def\svgwidth{0.85\columnwidth}
        \small\quad\quad
        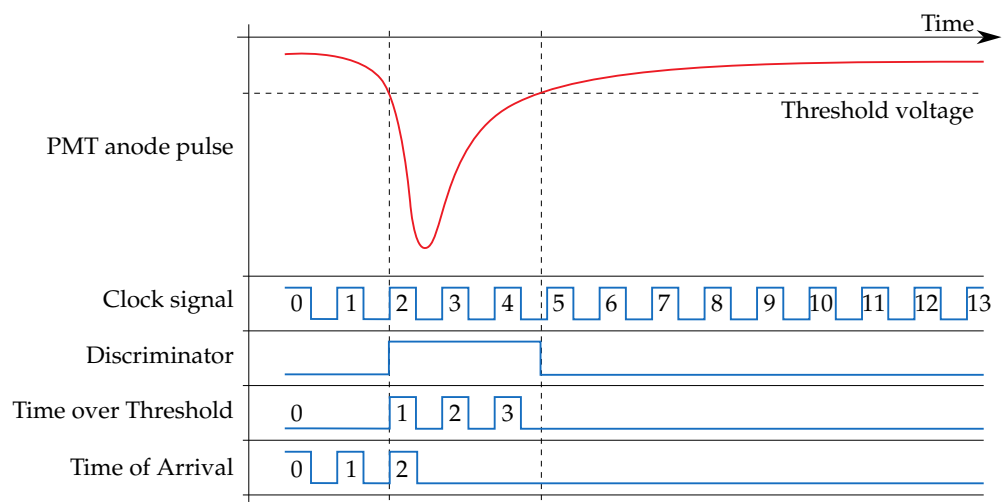
    }
    \caption{Digitization of an analog pulse using Time over Threshold~(ToT) and Time of Arrival~(ToA). A discriminator compares analog pulse~(red, not to scale) is to an adjustable threshold voltage. While ToT~counts the number of clock cycles spent in excess of the threshold, ToA~latches the latest timestamp when threshold is exceeded for the first time. In this example, the output ToA~is~2 and ToT~is~3.}
    \label{fig:daq:nikhef_poms:tot}
\end{figure}

\coco was configured by a second~ASIC called \promis~(PMT Read Out Mixed Signal)~\cite{gajanana2013asic}, which acted as a main point of communication for a single PMT. \promis~responded to slow control commands over \isquaredc~bus and produced data as low-voltage differential signal~(LVDS). Internally \promis generated two voltages, adjustable with 8-bit~depth: a high voltage control signal that was routed to \coco, and a threshold voltage that controlled trigger sensitivity. \promis~used a self-triggering mechanism that continuously compared amplified analog pulse from PMT~anode to the configured threshold voltage, and issued triggers whenever pulse amplitude exceeded threshold for the first time. Incoming trigger initiated pulse digitization, which was performed by counting clock cycles until the amplitude of the pulse dropped below the threshold voltage, as illustrated in~\Cref{fig:daq:nikhef_poms:tot}. This gave rise to a quantity known as Time over Threshold~(ToT), which was registered with time resolution below~\qty[round-precision=0]{2}{\nano\second} and represented a single PMT~hit. The shortest supported time between two successive hits was~\qty[round-precision=0]{25}{\nano\second}, which could be viewed as dead~time of this technology.

\promis encoded ToT~data into pulse-width modulated~(PWM) LVDS~signals, and transmitted them to a Calamares~board~\cite{real2016digital}. Since \promis~was located in a PMT~base and Calamares resided several meters away in a container in the middle of the POM, connection between the two was facilitated by a standard Category~5~(\cat5) twisted-pair cable~\cite{isoiec11801}. Due to their abundant use in modern networking and telecommunications, these cables were easy to handle, inexpensive and readily available in large quantities. Succeeding an older~PCB called Octopus, Calamares acted as a hub for the entire POM. With a distinctive elongated form factor, a pair of Calamares boards was able to accept up to 30~RJ45 connections from PMT~bases. Their main responsibility was to distribute power, clock signals and multiplex \isquaredc~communication.

\begin{figure}
    \centering
    \input{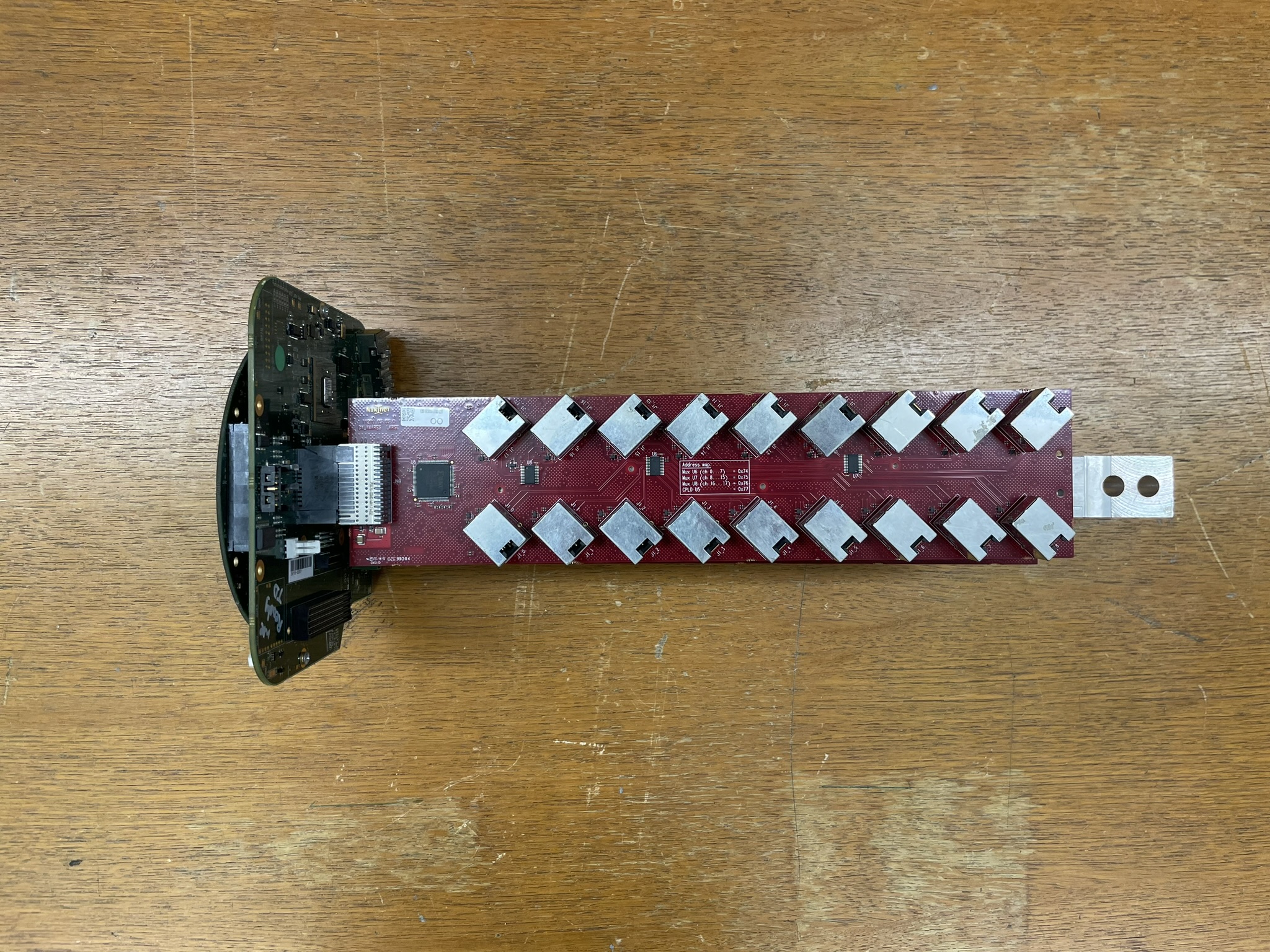}
    \caption{A CLB with a pair of Calamares boards. Note that one Calamares board obscures another board on the underside. Upstream and downstream ports are annotated.}
    \label{fig:daq:nikhef_poms:clb_calamares}
\end{figure}

As \Cref{fig:daq:nikhef_poms:clb_calamares} shows, Calamares interfaced with a Central Logic Board~(CLB)~\cite{real2016digital}, which acted as the main point of contact for the POM. CLB~was primarily responsible for aggregating measurements from all~PMTs of the POM. For each channel, it decoded signals from~\promis and timestamped them with a \qty[round-precision=0]{10}{\mega\hertz}~clock signal, complementing hit ToT with a quantity called Time of Arrival~(ToA). To maintain synchronization with the rest of the detector, the field-programmable gate array~(FPGA) inside~CLB continuously disciplined its on-board oscillator using methods described in \Cref{sec:daq:timing}. In addition to processing PMT~hits, CLB controlled a flasher circuit known as Nanobeacon~\cite{aiello2022nanobeacon}, which was capable of driving a light-emitting diode~(LED) in constant phase with timestamping clock. Nanobeacon delivered light pulses with tunable frequency and intensity, allowing~PMTs to be precisely calibrated inside the detector following its deployment. Upstream communication from a~CLB was facilitated by a fibre optic \qty[round-precision=0]{1}{\giga\bit}~Ethernet link towards fanout electronics containers, which forwarded data to \daq~computers.

\begin{figure}
    \centering
    \includegraphics[width=\columnwidth,trim={0 250pt 0 50pt},clip]{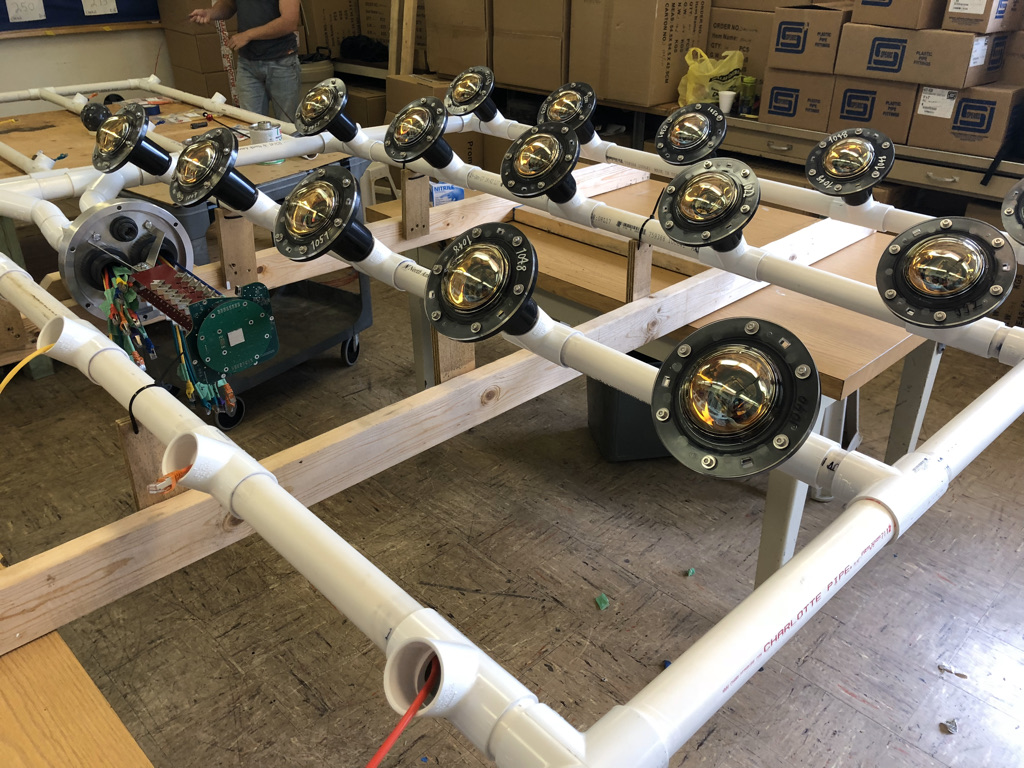}
    \caption{Partially constructed Nikhef POM with open electronics container, exposing a~CLB with Calamares~boards. Photograph courtesy of S.~Bash.}
    \label{fig:daq:nikhef_poms:plane_with_container}
\end{figure}

\begin{figure}
    \centering
    \includegraphics[width=\columnwidth,trim={0 170pt 0 90pt},clip]{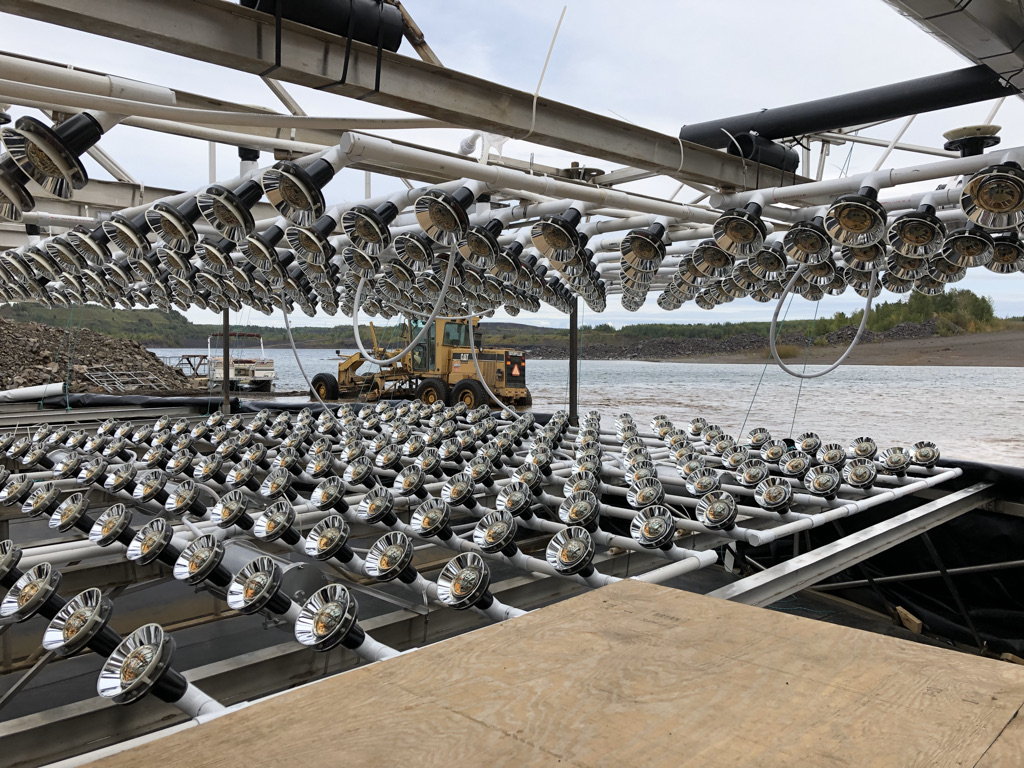}
    \caption{Fully assembled Nikhef POMs mounted on \chipsfive~endcaps. On the top endcap, veto~PMTs are oriented up. The rest are oriented inward at~\ang[round-precision=0]{45} facing the \numi~beam to maximize light collection. Photograph courtesy of S.~Bash.}
    \label{fig:daq:nikhef_poms:plane}
\end{figure}

CLB together with Calamares boards were installed in a cylindrical pressure-resistant container manufactured from aluminum, which is shown in \Cref{fig:daq:nikhef_poms:plane_with_container}. This compartment received downstream \cat5 cables from PMTs and upstream data fibres and power lines from higher levels of the \daq~system. Besides PCBs the container also carried a small power supply~(PSU) that stepped~\qty[round-precision=0]{240}{\volt} down to~\qty[round-precision=0]{12}{\volt} for POM electronics. To prevent water ingress into the container, all openings were fitted with waterproof feedthroughs. Similar features were installed throughout the POM near each~PMT, dividing pipes into compartments that could independently resist flooding. \Cref{fig:daq:nikhef_poms:plane} shows fully assembled Nikhef POMs installed in \chipsfive.

\subsection{Madison POMs} 
\label{sec:daq:madison_poms} 

POMs of the Madison type in many aspects mirrored their Nikhef counterparts. For instance, they used the same building elements based on PVC~tubing and had approximately similar form factor. Since they were seen as testbeds for the next generation of \daq~electronics, they were only deployed to the bottom endcap of \chipsfive and their numbers were relatively limited until thorough commissioning could be conducted. In the initial phase of deployment, the detector was instrumented with \num[round-precision=0]{30}~Madison POMs carrying \num[round-precision=0]{450}~PMTs in total. A single Madison POM contained \num[round-precision=0]{15}~Hamamatsu~PMTs, which were driven by novel electronics developed by \chips in collaboration with WIPAC~\cite{huber2018icetop}. For consistency, these components were organized to match the architecture and data flow of Nikhef hardware. PMT bases were fitted with microcontrollers, which transmitted measurements to a~PCB installed in a cylindrical container. From there data travelled upstream towards fanouts, which forwarded it to computers on the shore.

\begin{figure}
    \centering
    \input{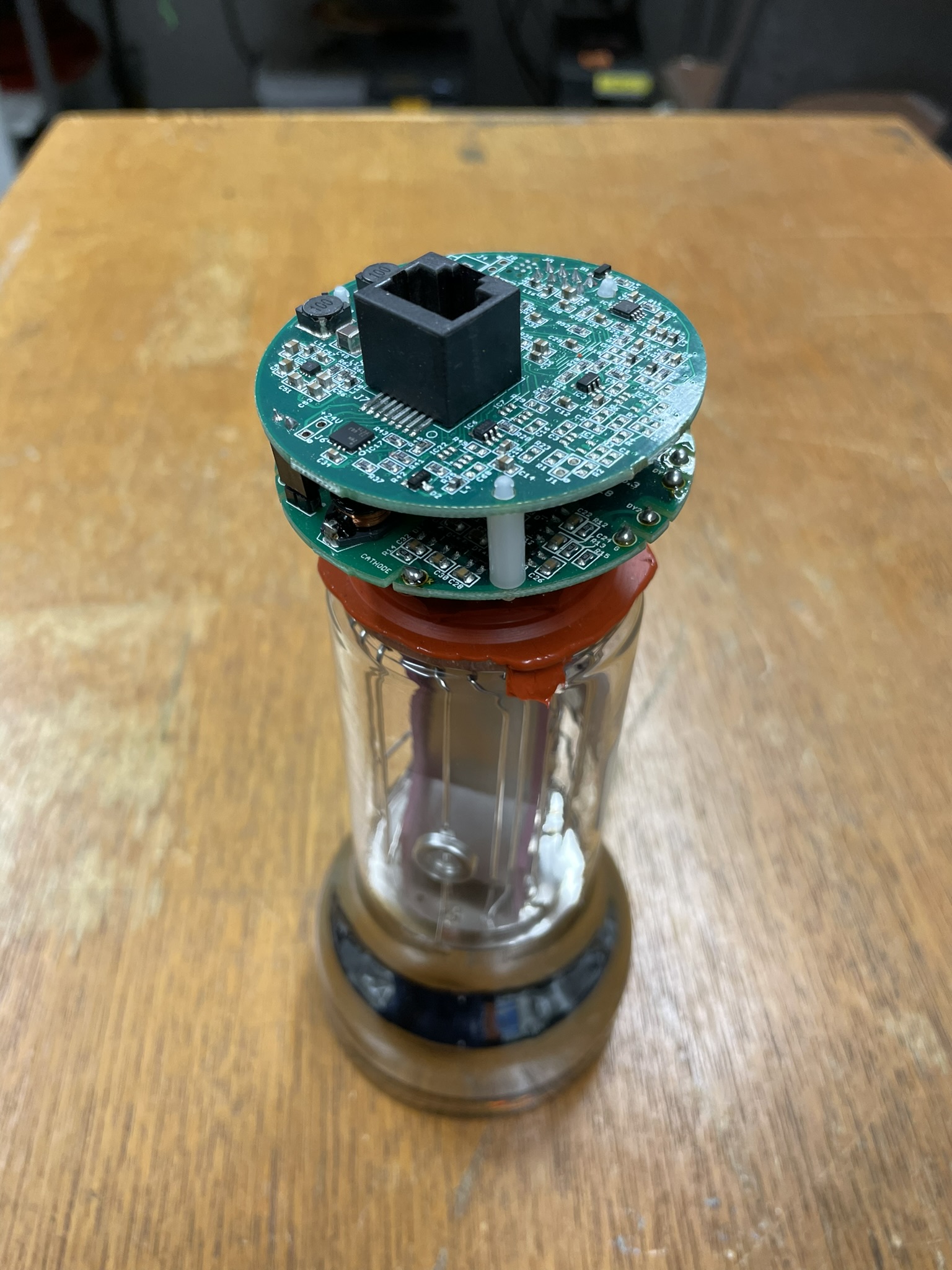}
    \caption{High voltage base of a Hamamatsu~PMT. The upper~PCB is the \microdaq, which is responsible for analog pulse digitization and high voltage control. The lower~PCB implements CW~voltage multiplier circuit that drives the~PMT.}
    \label{fig:daq:madison_poms:microdaq}
\end{figure}

\Cref{fig:daq:madison_poms:microdaq} shows the base of a Hamamatsu~PMT, which contains a microcontroller circuit called~\microdaq. This component was the main point of communication for the~PMT, and can be functionally viewed as a combination of \coco and \promis. Just like \coco, \microdaq drove a CW~voltage multiplier to generate high voltage for the~PMT. And just like \promis, \microdaq amplified and digitized analog pulses from PMT~anode. With that said, \microdaq implemented several desirable improvements over its counterparts.

Firstly, given its superior computing power the on-board STM32~microcontroller was capable of decoding time signals, which were supplied in the IRIG-B format~\cite{irig2000}. This allowed it to assign ToA simultaneously with ToT to each registered hit. In contrast, \promis was able to only measure ToT that was later timestamped by a~CLB several meters away, subject to cable delay. Secondly, in addition to self-triggering \microdaq implemented a CPU~trigger that periodically sampled analog pulse regardless of whether threshold was exceeded or not. This permitted a much more accurate characterization of backgrounds. Finally, even though \microdaq internally used a~\qty[round-precision=0]{180}{\mega\hertz} clock (corresponds to \qty[round-precision=1]{5.5}{\nano\second}~period), through resourceful engineering it delivered sub-nanosecond time resolution. This was achieved by routing the analog pulse through a tapped delay line~(TDL) with 8~taps, which were latched into registers upon triggering. By comparing contents of these registers, a firmware determined the precise time when threshold was exceeded with resolution of~\qty[round-precision=4]{0.6875}{\nano\second} without requiring expensive components such as~FPGAs or time-to-digital converters~(TDCs). This way, \microdaq{}s were seen as precision \daq~hardware, which could be manufactured at scale for a price of a conventional PCB, consistent with \chips~methodology.

\microdaq accumulated measurements in a volatile hit buffer that was divided into 32~pages. Under normal operation, contents of these pages were extracted by a~PCB known as Badger board, which was located inside a dedicated plastic container in the middle of the POM. Similarly to Nikhef instrumentation, connection between these two elements was facilitated by standard \cat5~cabling that carried LVDS~communications over distances of several meters. However, here data was transmitted over standard~RS-232 serial link~\cite{rs232}.

\begin{figure}
    \centering
    \input{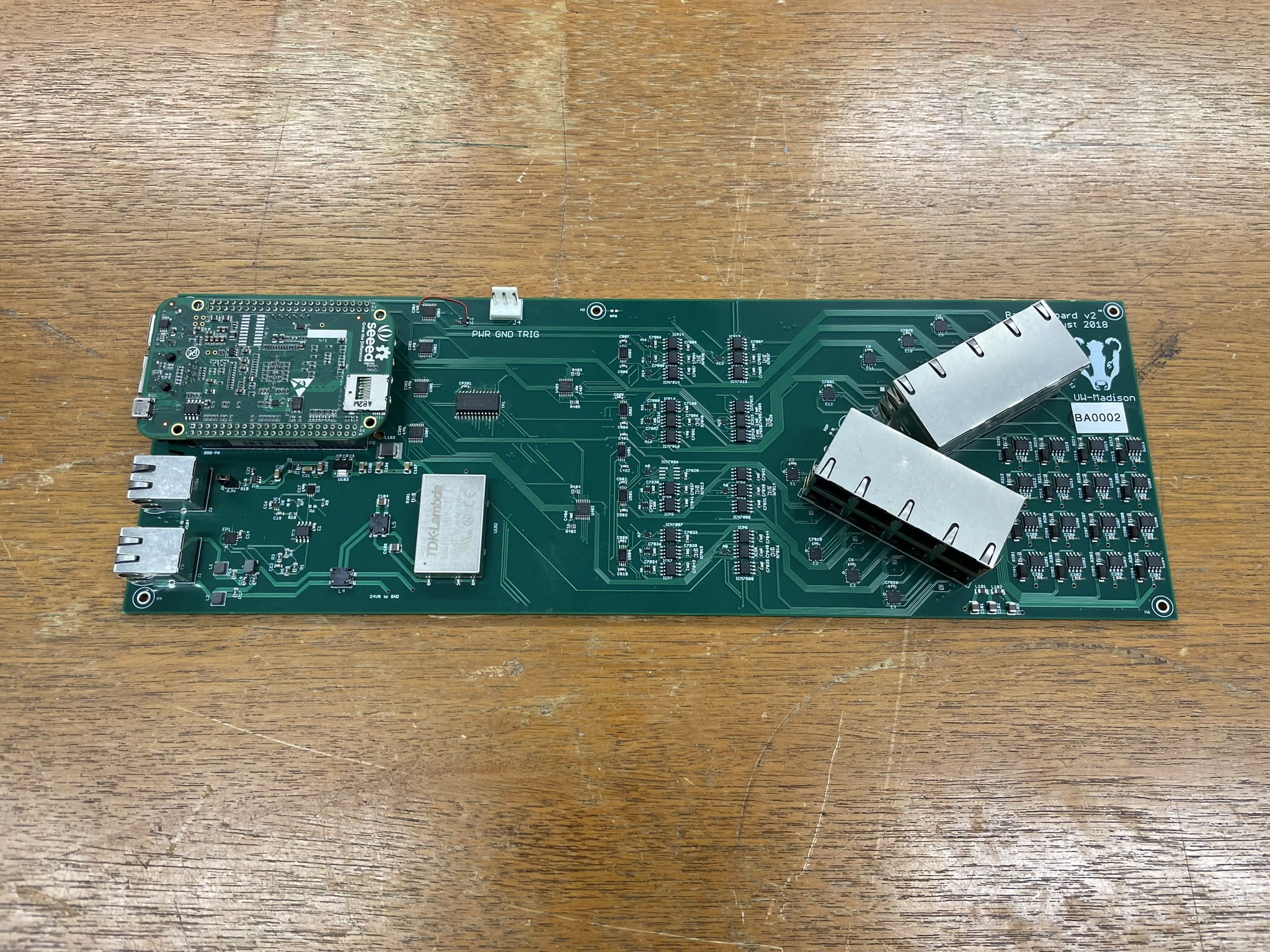}
    \caption{Badger board with \bbg SBC. Upstream and downstream ports are annotated.}
    \label{fig:daq:madison_poms:badgerboard}
\end{figure}

A Badger board is shown in \Cref{fig:daq:madison_poms:badgerboard}. Much like Calamares boards with a CLB, this~PCB was designed to aggregate their measurements from all PMTs of the POM, and  distribute power and timing signals. To this end it incorporated a single-board computer~(SBC) called SeeedStudio \bbg (abbreviated as BeagleBone)~\cite{seeed2024bbg}, which interfaced with the PCB by a pair of 46-pin~headers that routed \isquaredc bus and data links to its general-purpose input/output~(GPIO) pins. The BeagleBone ran a customized version of the Linux operating system, which was hosted from a microSD~card. In this environment, a suite of C~programs known as Field~Hub App controlled and read out up to 16~\microdaq{}s.

Since hits were already timestamped upon digitization, their processing was considerably simplified. Specifically, the BeagleBone did not need to decode timing signals, which freed up its computing power for preliminary analysis and quality control. Communications upstream were implemented over metallic \qty[round-precision=0]{100}{\mega\bit}~Ethernet link. This channel was also used for monitoring and maintenance interventions, which typically entailed remote access over Secure Shell Protocol~(SSH).

\begin{figure}
    \centering
    \includegraphics[width=\columnwidth,trim={0 220pt 0 120pt},clip]{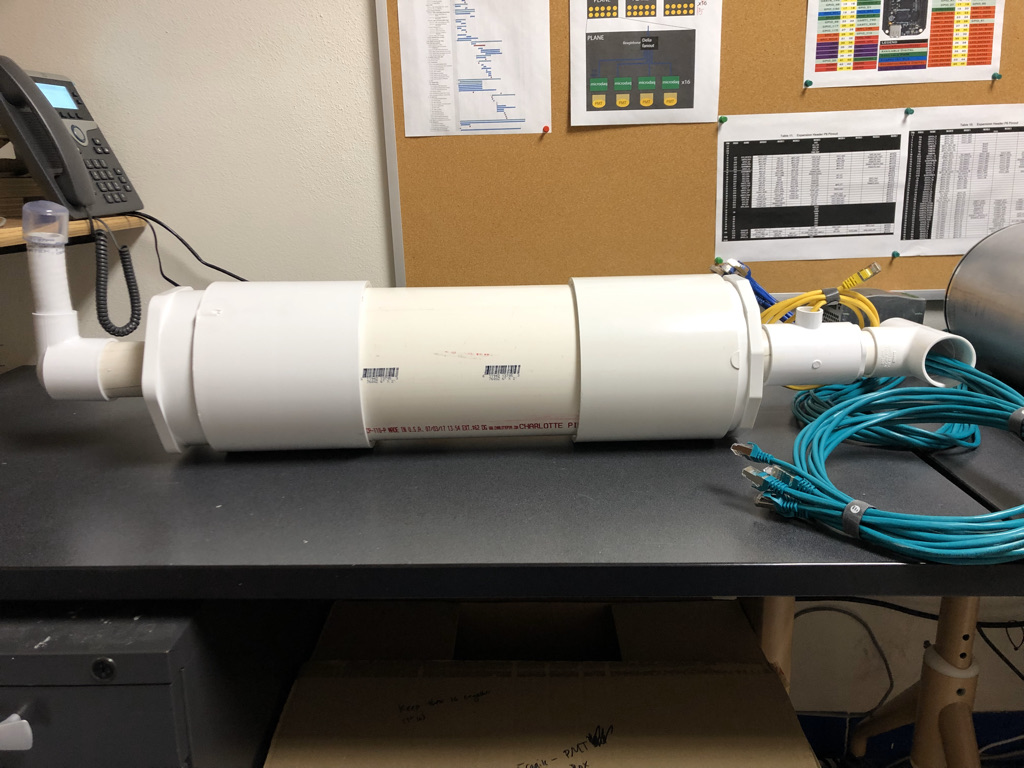}
    \caption{Sealed electronics container of a Madison POM. Inside is a Badger board~(middle compartment) with a Nanobeacon~(capped pipe on the left). \cat5~cabling enters the container from the right side. Photograph courtesy of S.~Bash.}
    \label{fig:daq:madison_poms:container}
\end{figure}

Just like CLBs, Badger boards were able to drive a single Nanobeacon flasher in synchronization with timestamping clock for the purposes of PMT~calibration. Intensity and frequency of delivered pulses were programmed from the BeagleBone over \isquaredc~bus. The entire electronics assembly consisting of a single Badger board, a Nanobeacon and a \qty[round-precision=0]{24}{\volt}~PSU was enclosed in a plastic cylinder, modelled after its aluminum counterpart from Nikhef POMs. This design, visible in \Cref{fig:daq:madison_poms:container}, aimed to reduce costs by employing standard PVC~plumbing parts with glued joints instead of machined components with feedthroughs.

\begin{figure}
    \centering
    \includegraphics[width=\columnwidth,trim={190pt 0 0 200pt},clip]{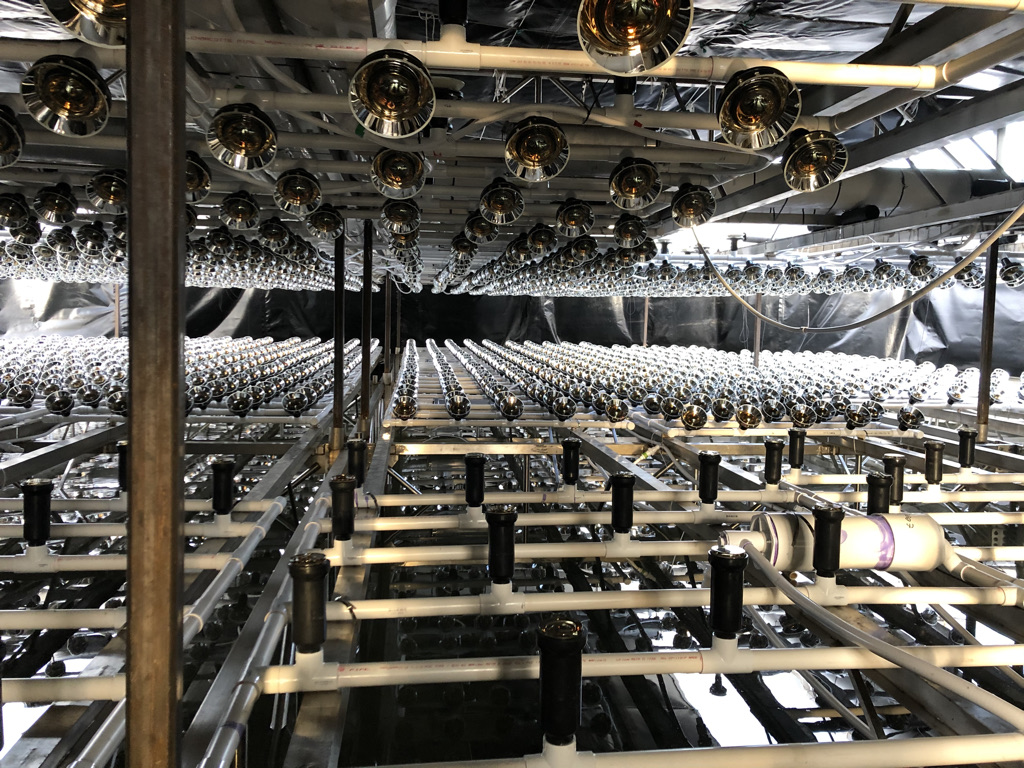}
    \caption{Fully assembled Madison POMs~(front) mounted alongside their Nikhef counterparts~(back) on \chipsfive~endcaps. Note that PMT~density of Madison POMs is half that of Nikhef POMs. Photograph courtesy of S.~Bash.}
    \label{fig:daq:madison_poms:plane}
\end{figure}

Madison POMs were waterproofed in similar fashion to Nikhef POMs. PVC~tubing was divided by bulkheads into a series of compartments, which were able to resist flooding independently. Thanks to material consistency, PVC~glue was generously used along all joints and interfaces. The end result is presented in \Cref{fig:daq:madison_poms:plane}, which shows fully assembled Madison~POMs installed in \chipsfive.

\section{Infrastructure hardware} 
\label{sec:daq:infra-hardware} 

Besides POMs, which represented the bulk of submerged instrumentation, the \chipsfive detector also included auxiliary \daq~subsystems that supplied all elements with power, accurate timing and high-speed network links. This equipment represented the infrastructure that interconnected submerged devices and patched them to a hut on the shore of the mining pit, where additional hardware was installed. This included mainly computers, network switches, power supplies, relays and timing equipment.

\subsection{Time distribution} 
\label{sec:daq:timing} 

To ensure that flashes of Cherenkov light were seen as coincident from multiple vantage points, POM instrumentation needed to timestamp PMT~hits with a clock that was precisely synchronized to the same reference across the entire detector. Furthermore, given \chipsfive~dimensions it was critical for this synchronization to be accurate within at least units of nanoseconds. This motivated use of White Rabbit~(WR) timing system~\cite{lipinski2011white}.

Originally developed by the European Organization for Nuclear Research~(CERN) between 2008 and~2012, WR~is an ongoing research project aiming to deliver sub-nanosecond time synchronization for applications in physics research and industry. At its inception, WR~protocol specification as well as its core implementation were made openly available in the public domain, which allowed researchers as well as private companies to freely adopt this technology. Since then many manufacturers have expanded their portfolios to include WR-compatible products. In~2019, thanks to widespread use of WR the protocol was incorporated into the IEEE~1588 standard as the high accuracy profile of Precision Time Protocol~(PTP) version~3~\cite{ieee1588}.

From user perspective, a great advantage of~WR is its elegant use of abundant network infrastructure. Instead of requiring specialized cables, hubs or receptacles, the timing system is implemented as an extension of a conventional \qty[round-precision=0]{1}{\giga\bit}~Ethernet network carried over single-mode bidirectional optical fibres. This makes~WR much more interoperable, as it permits WR-enabled peripherals, also called \textit{WR~nodes}, to readily communicate with any other devices regardless of their compatibility. Support for the WR~protocol is detected automatically during link negotiation phase and enabled only if implemented by both sides. When enabled, network interface controllers~(NICs) of WR~nodes regularly exchange time synchronization packets injected into ordinary Ethernet traffic. Since these data frames are immediately consumed by WR~hardware on both ends of the link, they are rendered invisible to any software supervising the data exchange. This is highly desirable, as it allows a single fibre to simultaneously carry communications as well as precise time.

While low-level implementation of~WR is beyond the scope of this text, its basic operating principle is not difficult to summarize. Both participants of a WR~link have oscillators, which can be viewed as local clocks. In their initial state these clocks would inherently drift apart given a sufficient period of time. To prevent this, nodes on both sides exchange messages that establish a phase-locked loop~(PLL). This gives rise to a hierarchy where one node, known as the \textit{slave}, perpetually adjusts its clock to match the other, known as the \textit{master}. Since a~PLL maintains constant but not necessarily zero phase between clocks, additional messages are continuously exchanged to track and subtract this phase offset. Finally, to counteract delays due to fibre length WR~packets are timestamped by their senders, allowing both nodes to measure and subtract roundtrip latency. Accuracy of this measurement can be further improved through a specialized calibration procedure that probes asymmetry of transmit/receive operations of the specific model of small form-factor pluggable transceiver~(SFP). When all these approaches are applied in concert, two directly connected WR~nodes can typically synchronize their local clocks to within~\qty[round-precision=0]{100}{\pico\second} over potentially vast physical distances. For instance, in~2021 a WR-based system was able to synchronize clocks between New~Jersey and Chicago over the distance of~\qty[round-precision=0]{1350}{\kilo\meter} with mean offset of~\qty[round-precision=0]{112 +- 139}{\pico\second}~\cite{safran2024long}.

\begin{figure}
    \centering
    \includegraphics[width=0.8\columnwidth]{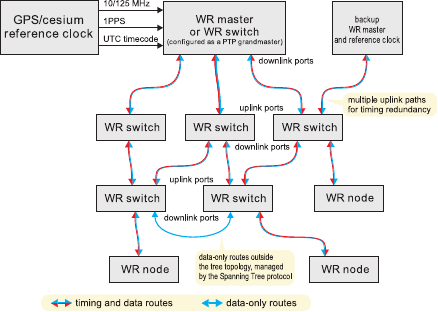}
    \caption{A typical WR~network. Time is distributed from a reference clock through a topology of switches down to nodes. Blue-red arrows indicate fibre links, which transmit data as well as timing signals. Diagram taken from reference~\cite{lipinski2011white}.}
    \label{fig:daq:timing:wr_network}
\end{figure}

When WR~is deployed over larger networks, the same time synchronization process is repeated for each adjacent pair of WR~nodes, forming a chain of mutually synchronized devices. This introduces one practical drawback of this technology: it is usually undesirable to introduce incompatible elements to higher levels of WR~networks, as they strip away timing information and prevent it from propagating. Since conventional Ethernet~switches would have this exact effect, WR~offers its own equivalent network element, known as \textit{WR~switch}, which acts as a WR~node on all ports~\cite{wrs2020}. WR~switches allow WR~networks to branch into tree-shaped topologies, where switches at each level act as masters for the level below and slaves for the level above. In addition, these topologies allow for improved reliability where multiple clock sources can offer redundant failover capacity in case of faults. The WR~switch situated in the root of the tree is called the \textit{grandmaster}, since it distributes its clock to all other nodes in the system. This is topology illustrated in \Cref{fig:daq:timing:wr_network}.

Grandmasters typically keep their own free-running clocks or receive clock reference from external time sources, such as atomic clocks or Global Navigation Satellite System~(GNSS) receivers. WR~switch hardware expects this information to be supplied as a pair of standardized timing signals:%

\begin{enumerate}
    \item
    \qty[round-precision=0]{10}{\mega\hertz} sine wave, which precisely marks time within the scope of a single second.

    \item
    IRIG-B square wave signal of frequency \qty[round-precision=0]{1}{\hertz}\footnote{IRIG-B standard requires complete ToD~information to be transmitted every second. However, bits representing this data are sent at~\qty[round-precision=0]{100}{\hertz}. \cite{irig2000}}, which marks the turnover of a second and encodes time-of-day~(ToD) information.
\end{enumerate}

This method represents an attractive option for physics projects such as \chipsfive, which not only needed to synchronize instrumentation within the scope of the detector, but also aimed to correlate PMT~timestamps with spills of the Fermilab accelerator hundreds of kilometres away. Without necessitating a dedicated communications line connecting the two facilities, GNSS~receivers could be used on both ends of the baseline to retrieve International Atomic Time~(TAI), which was used as a common clock reference. This was fed into grandmaster WR~switches that synchronized the rest of equipment on site.

\subsection{Networking and power infrastructure} 
\label{sec:daq:network_power_infra} 

\begin{figure}
    \centering
    \includegraphics[width=\columnwidth]{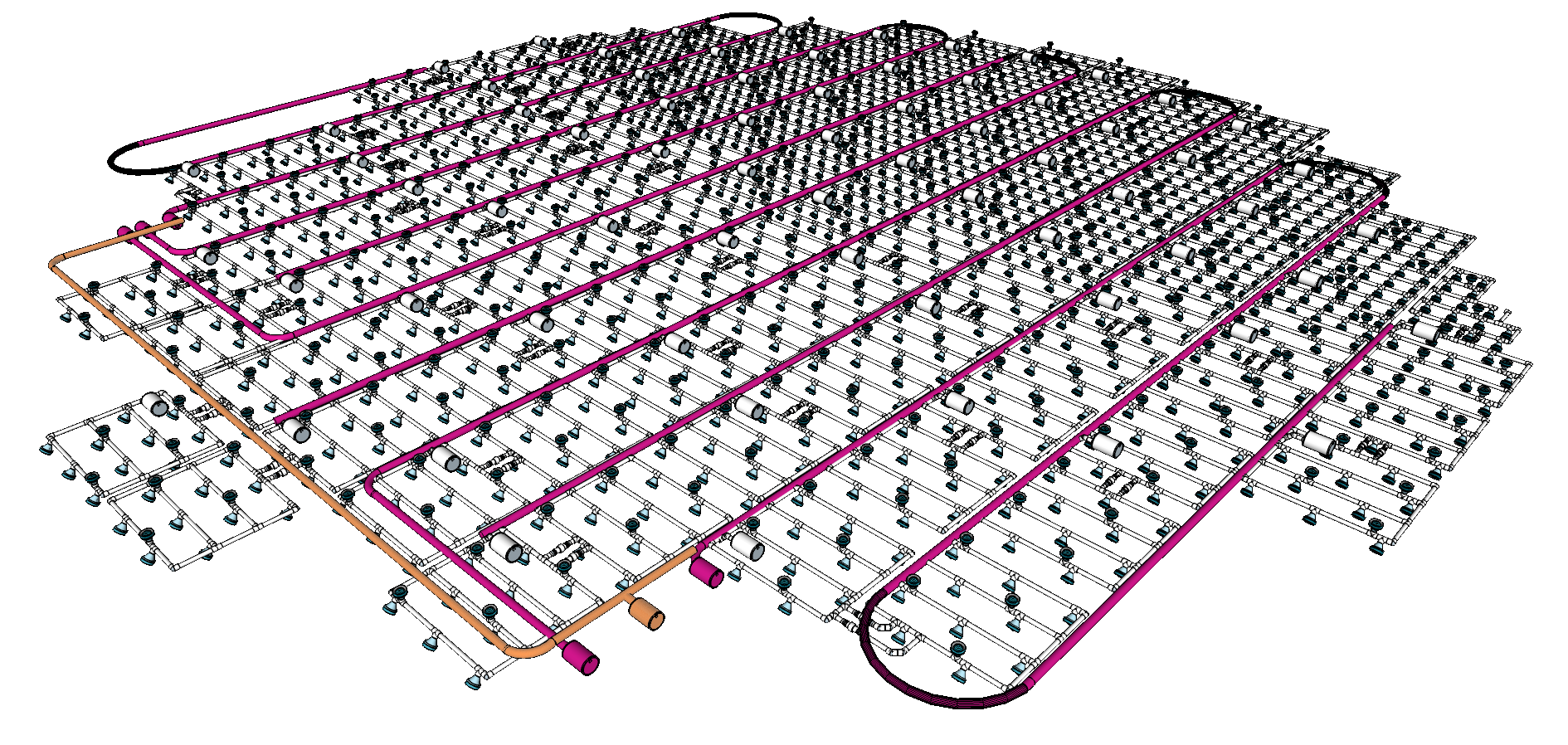}
    \caption{Rendering of the \chipsfive top endcap. Nikhef POMs~(white) are interconnected by manifolds (orange, pink). Picture courtesy of T.~Dodwell.}
    \label{fig:daq:network_power_infra:manifolds}
\end{figure}

At the \chipsfive site in Minnesota, WR~was deployed as a part of broader infrastructure that distributed power and communications from shore equipment to submerged~PMTs. This system consisted of several cylindrical containers, which structurally resembled those carried by POMs. While most of them were so-called \textit{fanouts}, which serviced groups of POMs, one container, known as the \textit{junction box}, was the master hub that was patched directly to the umbilical. Interconnections between the junction box, fanouts and POMs were facilitated by elaborate conduits called \textit{manifolds}, which protected cables and optical fibres inside waterproof PVC~hoses. These are illustrated in \Cref{fig:daq:network_power_infra:manifolds}.

Fanouts were responsible for power switching and multiplexing WR~network towards a group of POMs. Since their technology depended on POMs, there existed two types of fanouts that aligned with POM types. In its initial phase of deployment, \chipsfive was equipped with 5~Nikhef fanouts and 1~Madison fanout. 

Nikhef fanouts contained a pair of WR~switches, which were cross-linked for redundancy. In the upstream direction, these switches were patched to shore huts through the junction box. In the downstream direction, they interacted directly with~CLBs on-board Nikhef POMs. Since CLBs~were programmed as WR~slaves, they were able to access precise time as well as data links for~\daq and detector control. Besides WR~switches Nikhef fanouts also carried network-controlled relay boards, which interrupted power to individual POMs on remote command. To adapt metallic links from relay boards and WR~switch management interfaces to optical fibres, media converters were installed inside fanout containers.

\begin{figure}
    \centering
    \begin{subfigure}{0.45\columnwidth}
        \includegraphics[width=\columnwidth]{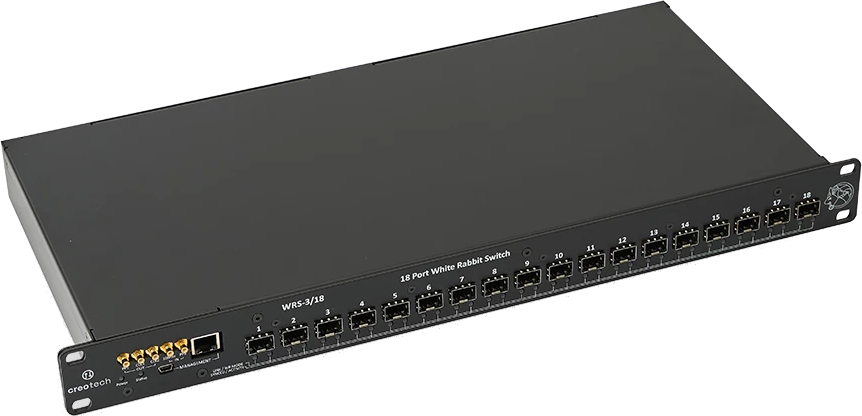}
        \caption{Commercially available WR~switch.~\cite{creotech2024wrswitch}\\Dimensions: \qtyproduct[round-precision=0]{447 x 44 x 223}{\milli\meter}}
        \label{fig:daq:network_power_infra:switch_commercial}
    \end{subfigure}%
    \quad%
    \begin{subfigure}{0.47\columnwidth}
        \centering
        \includegraphics[width=0.55\columnwidth]{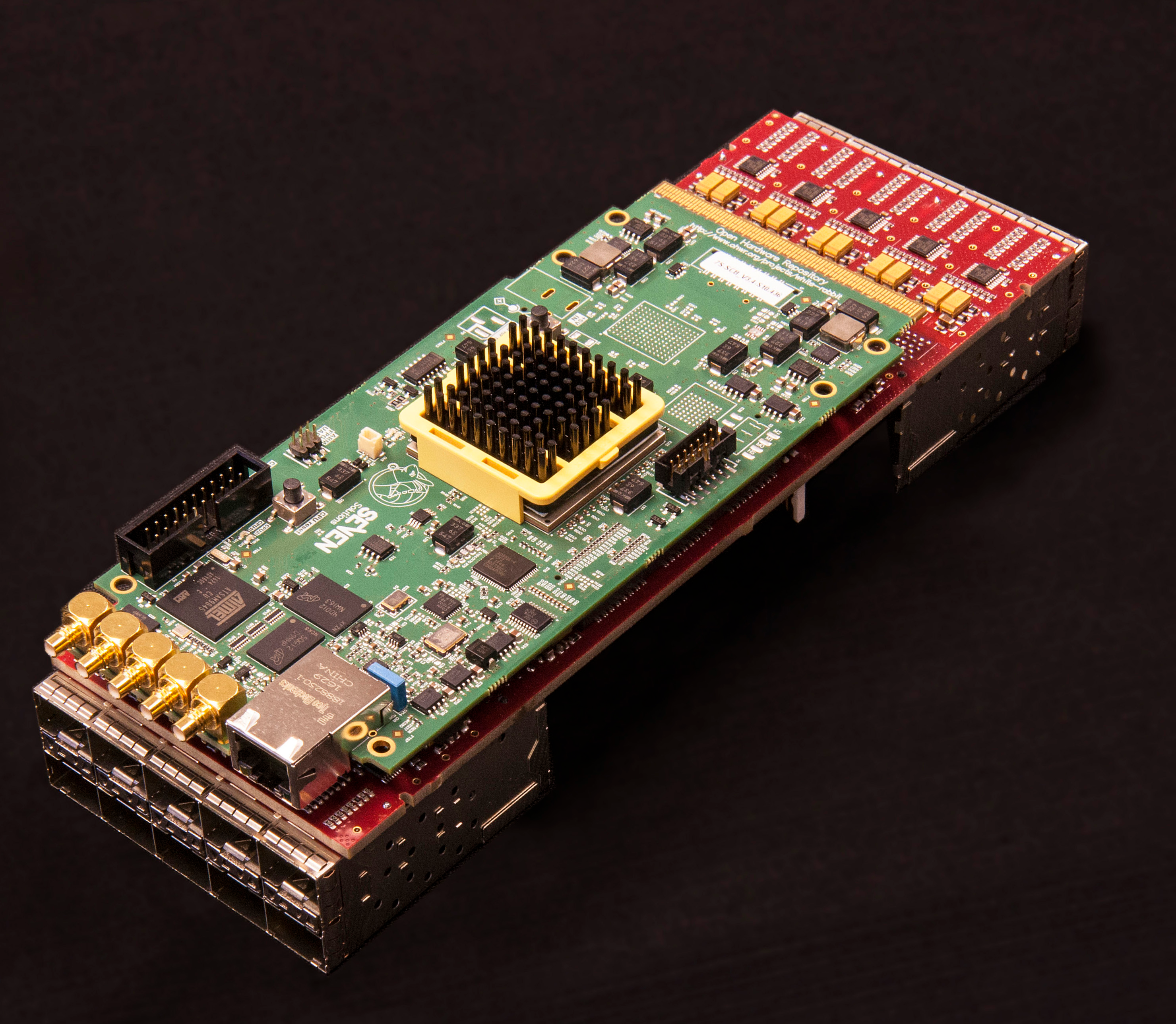} 
        \caption{Compact WR~switch used by \chipsfive.~\cite{wrchromium2020}\\Dimensions: \qtyproduct[round-precision=0]{75 x 47 x 236}{\milli\meter}}
        \label{fig:daq:network_power_infra:switch_compact}
    \end{subfigure}
    \caption{WR~switch comparison illustrating benefits of a compact~PCB developed by Nikhef~(b) over a commercially available solution~(a). Pictures are not to scale.}
    \label{fig:daq:network_power_infra:switch}
\end{figure}

During assembly of Nikhef fanouts, an interesting challenge was presented by large dimensions of WR~switches. Since these network elements were normally designed to operate in standard \qty[round-precision=0]{19}{\inch}~wide telecommunications racks, their commercially available enclosures were unsuitable for confined fanout containers. To overcome this issue and maximize packing efficiency, Nikhef developed a~PCB that was functionally equivalent to a WR~switch, but had considerably smaller footprint~\cite{wrchromium2020}. \Cref{fig:daq:network_power_infra:switch} shows how this device compares to a commercial WR~switch.

\begin{figure}
    \centering
    \input{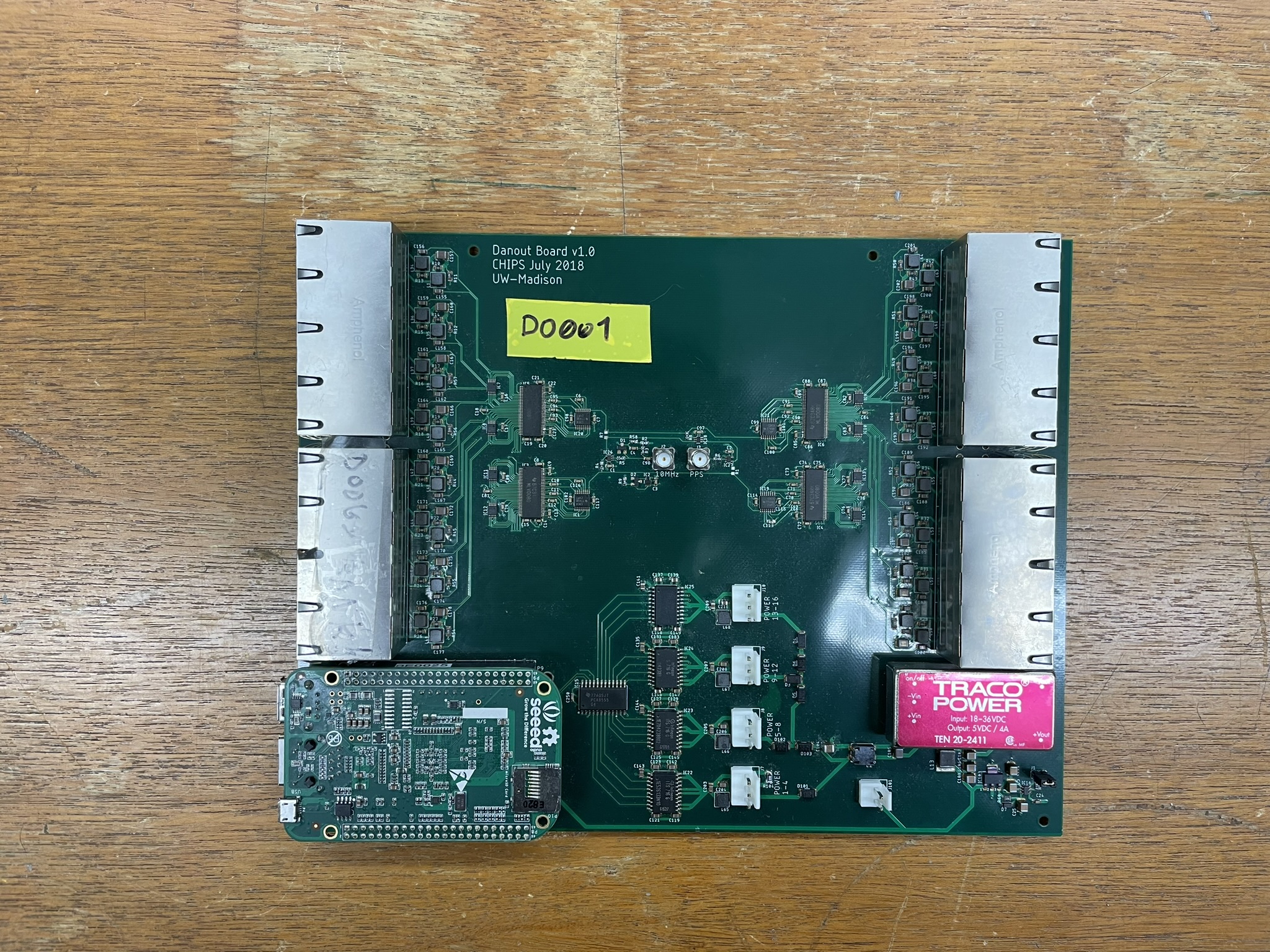}
    \caption{Danout board with \bbg SBC. Upstream and downstream ports are annotated.}
    \label{fig:daq:network_power_infra:danout_board}
\end{figure}

Much like their Nikhef counterparts, Madison fanouts were responsible for distributing communications and power to up to 16~Madison POMs. This was complicated by the fact that unlike~CLBs, Badger~boards were not directly compatible with the WR~standard. Madison fanouts therefore had to convert time from the WR~network to electrical timing signals, which were transmitted to POMs alongside data traffic. This was the purpose of a~PCB called Danout~board, which is shown in \Cref{fig:daq:network_power_infra:danout_board}. Danout board accepted \qty[round-precision=0]{100}{\mega\bit}~Ethernet links from upstream as well as a pair of timing signals, which it injected into a corresponding number of downstream links towards Badger~boards. Since these connections also carried power, the Danout~board was able to remotely switch individual POMs on or off through \isquaredc-controlled relays. The Danout~board was supervised by a BeagleBone~SBC, which was interfaced with the~PCB by the same header mechanism used in Badger~boards.

\begin{figure}
    \centering
    \begin{subfigure}{.45\columnwidth}
        \centering
        \includegraphics[height=4.7cm,trim={100pt 110pt 60pt 20pt},clip]{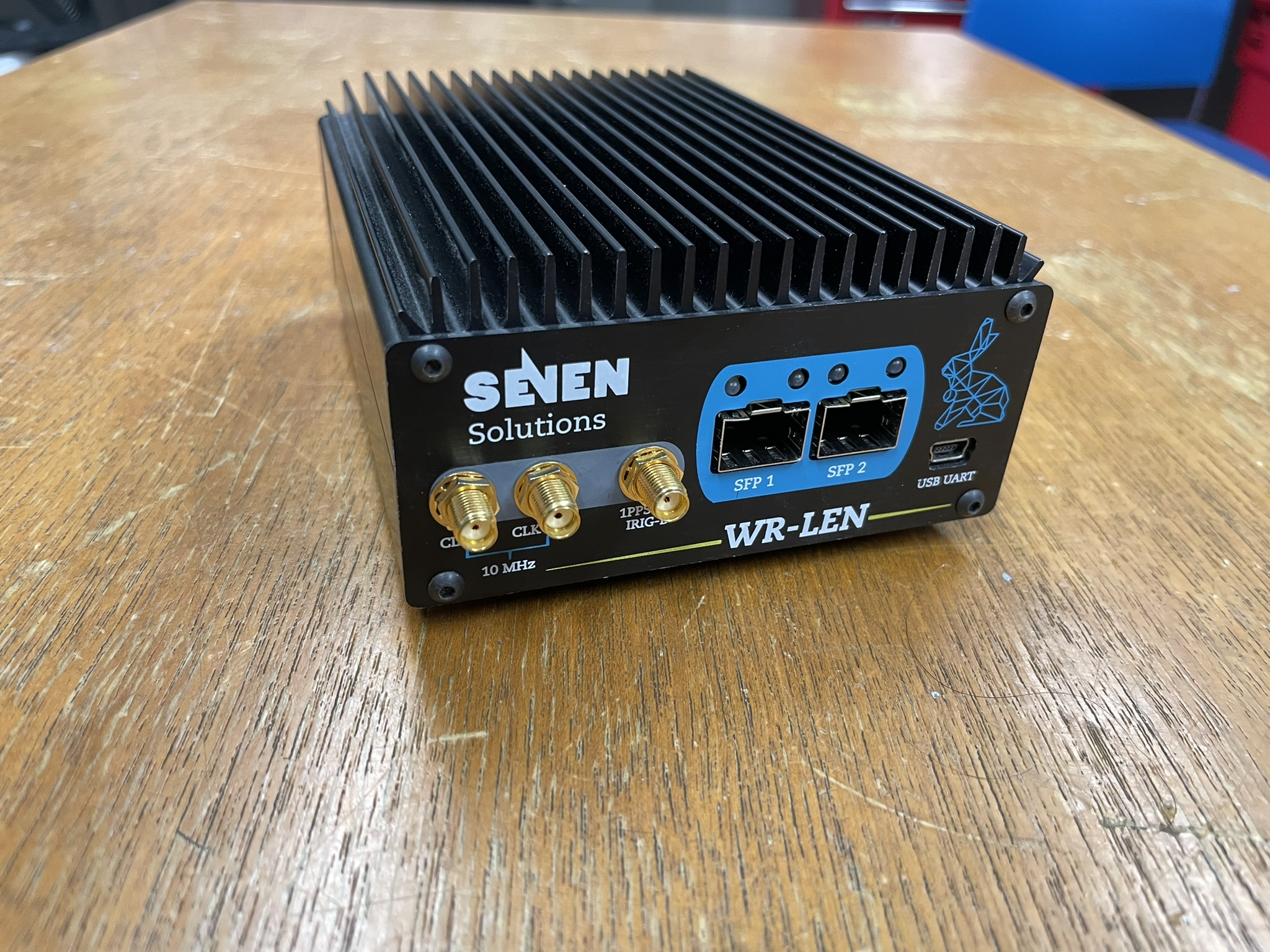}
        \caption{Front panel.}
        \label{fig:daq:network_power_infra:wrlen:front}
    \end{subfigure}%
    \qquad%
    \begin{subfigure}{.45\columnwidth}
        \centering
        \includegraphics[height=4.7cm,trim={60pt 50pt 40pt 20pt},clip]{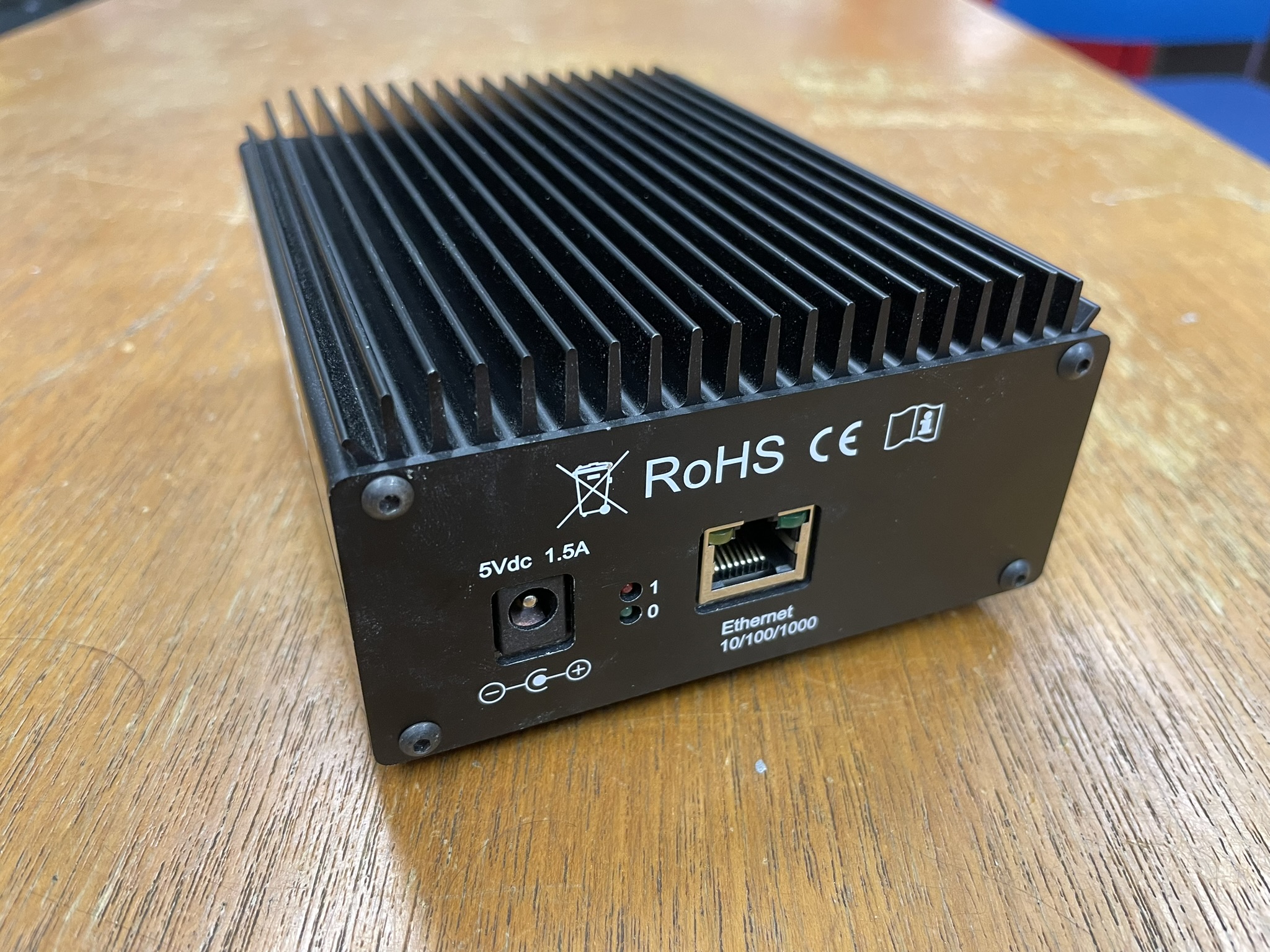}
        \caption{Back panel.}
        \label{fig:daq:network_power_infra:wrlen:back}
    \end{subfigure}
    \caption{Commercially available WR-LEN. The front panel shows timing signal outputs (circular SMA~ports on the left) and SFP~receptacles (rectangular ports in the middle) for bidirectional connection to a WR~network. The back panel has power connector~(left) and a single RJ45 port~(right) for \qty[round-precision=0]{1}{\giga\bit}~Ethernet link.}
    \label{fig:daq:network_power_infra:wrlen}
\end{figure}

Besides the Danout~board, Madison fanouts included conventional Ethernet~switches. These network elements multiplexed 16~lines of upstream data traffic into a single metallic~Ethernet line, which was routed through a device called \textit{WR~Lite Embedded Node}~(WR-LEN, shown in~\Cref{fig:daq:network_power_infra:wrlen})~\cite{sevensols2023wrlen}. As its name suggests, WR-LEN~was a WR~node programmed to act as a slave in the WR~network, which allowed it to receive precise time from higher levels of the \daq~system. WR-LEN~acted as a media converter and adapted metallic~Ethernet to a fibre optic link compliant with the WR~standard. In addition, WR-LEN~converted time from WR~network to~IRIG-B and \qty[round-precision=0]{10}{\mega\hertz}~clock signals, which were immediately consumed by the Danout~board. Just like Nikhef fanouts, the upstream fibre from~WR-LEN was patched to shore infrastructure through the junction box.

The junction box was the main hub for the entire submerged detector, which was responsible for distributing communications and power from the umbilical towards fanouts. To switch fanout power, the junction box carried two relay boards that were controlled remotely over the network, much like relays installed in Nikhef fanouts. For increased resilience against electrical faults, downstream power lines were protected by trip gates that automatically interrupted them if current surges were detected. In the event of flooding, this safety measure allowed the unaffected part of the detector to remain isolated from damaged subsystems.

WR~network from fanouts was transmitted through the umbilical by coarse wavelength division multiplexing~(CWDM). This method encoded traffic from 16~single-mode fibres into a single armoured multi-mode fibre that carried light of 32~distinct wavelengths along the umbilical. This removed the need for any Layer~2 switching inside the junction box. Instead, all WR~switches and~WR-LENs from fanouts interacted transparently with the network elements on the shore.

\begin{figure}
    \centering
    \input{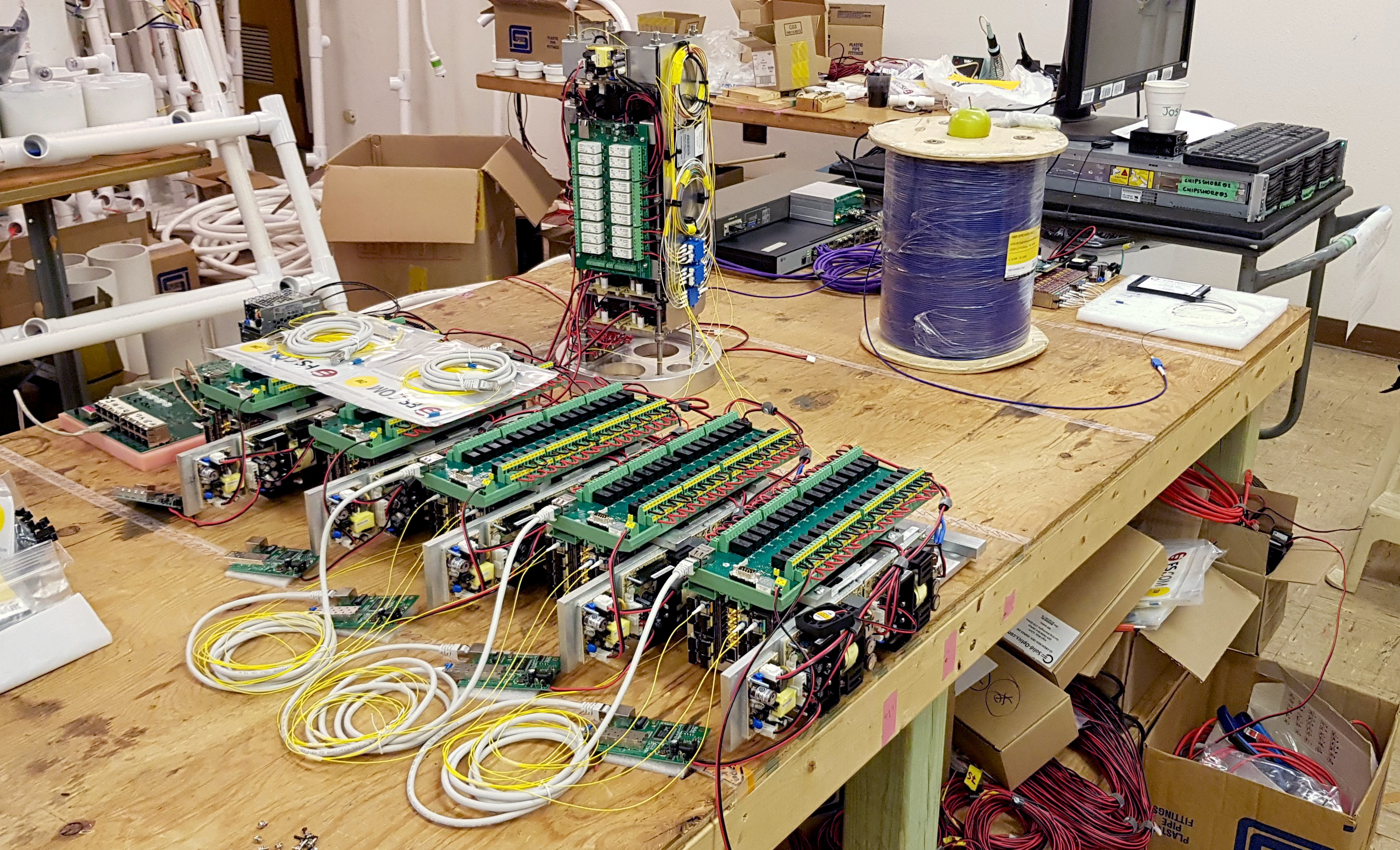}
    \caption{Infrastructure electronics before their installation into fanout containers and the junction box. Shore hut equipment is visible in the far right. Background photograph courtesy of J.~Tingey.}
    \label{fig:daq:network_power_infra:fanouts_junction_box}
\end{figure}

Since the junction box presented a single point of failure, extensive testing was performed prior to its installation in \chipsfive. The largest hazard was seen to be flooding due to a pressure seal fault. The design of the container mitigated this by employing feedthroughs and water-resistant bulkheads near all interfaces and openings, just like other electronics containers. To minimize probability of accidental failure following deployment, the junction box underwent dedicated pressure tests. In addition, all infrastructure electronics were jointly and independently tested prior to their installation in containers and waterproofing. \Cref{fig:daq:network_power_infra:fanouts_junction_box} shows one of these tests, providing an adequate overview of \daq~infrastructure deployed underwater.

\subsection{Shore facilities} 
\label{sec:daq:shore} 

\begin{figure}
    \centering
    \input{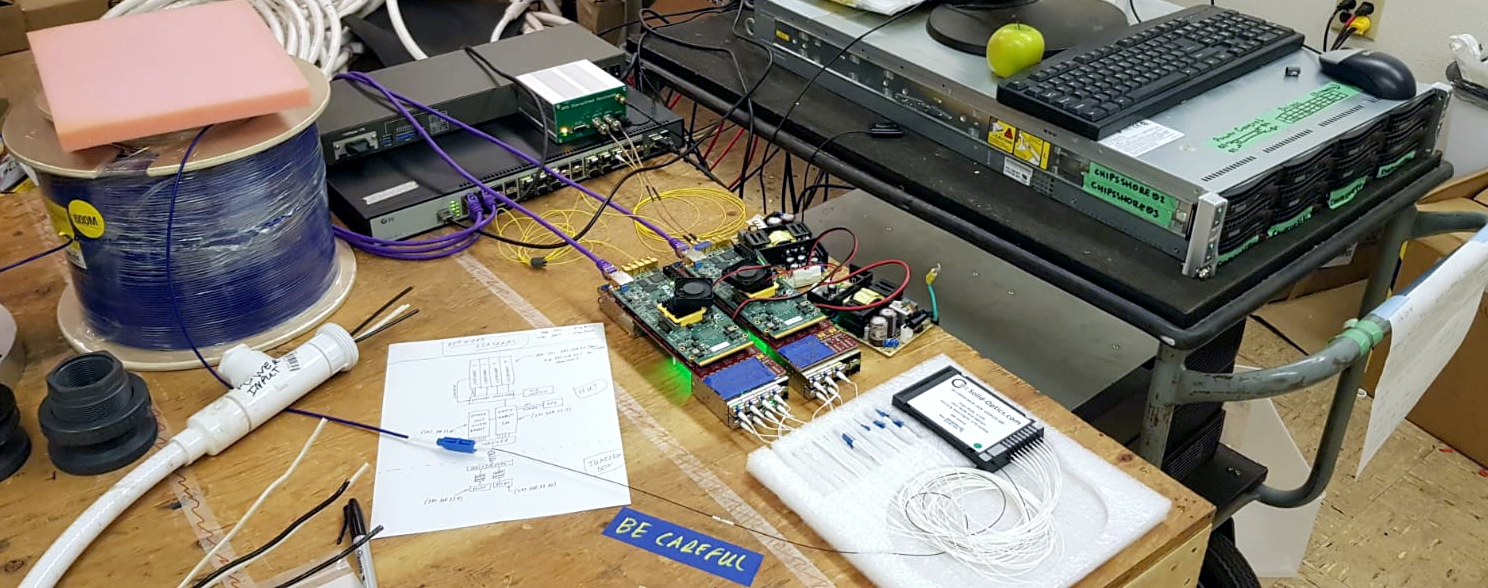}
    \caption{\daq infrastructure before installation in the electronics hut. Background photograph courtesy of J.~Tingey.}
    \label{fig:daq:shore:equipment}
\end{figure}

Power lines and the multi-mode fibre that emerged from the umbilical were routed into an electronics hut, which was situated on the shore of the mining pit alongside the water treatment hut. This small building housed remaining \daq~infrastructure that connected the deployed detector to power mains, remotely controlled its instrumentation and transmitted its measurements to Fermilab through the Internet. This equipment is shown in \Cref{fig:daq:shore:equipment}.

At the root of the WR~network was a pair of linked WR~switches, which consumed optical fibres from the umbilical through a CWDM~multiplexer. Contrary to intuition, these root WR~switches were not equivalent. One of them consumed timing signals from a GNSS~receiver, and therefore occupied the role of the grandmaster node in the time distribution hierarchy. In contrast, the other switch merely expanded the breadth of the network at the highest level. This way, WR~switches inside Nikhef fanouts and WR-LENs~inside Madison fanouts were always at most 2~hops away from the grandmaster.

To ensure that the WR~network was reachable from non-WR~elements installed upstream, root WR~switches were patched into a conventional \qty[round-precision=0]{10}{\giga\bit}~Ethernet switch. Since this switch was incompatible with WR, this divided the \daq~network into two bridged segments: a WR~segment that included both root WR~switches as well as all electronics beyond the umbilical, and a conventional Ethernet segment that contained peripherals dedicated to detector control and data handling. The most prominent of these devices were 4~computers that hosted \daq~software, which is further described in \Cref{sec:daq:software}. These computers primarily exchanged data and control messages with detector instrumentation, and were therefore adequately equipped to process incoming data in real-time with redundant capacity. In addition, they included sufficient persistent storage to hold logs and measurements for a period of at least several weeks.

The \daq~network was connected upstream to the Internet through pre-existing infrastructure managed by the PolyMet Mining Corporation. This permitted remote access for \daq~operators, hardware monitoring and alerting, data transfers towards Fermilab and low-latency delivery of \numi~spill triggers through \chipsfive~TDS. External access to the \daq~network was facilitated by encrypted virtual private network~(VPN) over Layer~2 Tunnelling Protocol~(L2TP).

\begin{figure}
    \centering
    \includegraphics[width=\columnwidth,trim={20pt 0 27pt 0},clip]{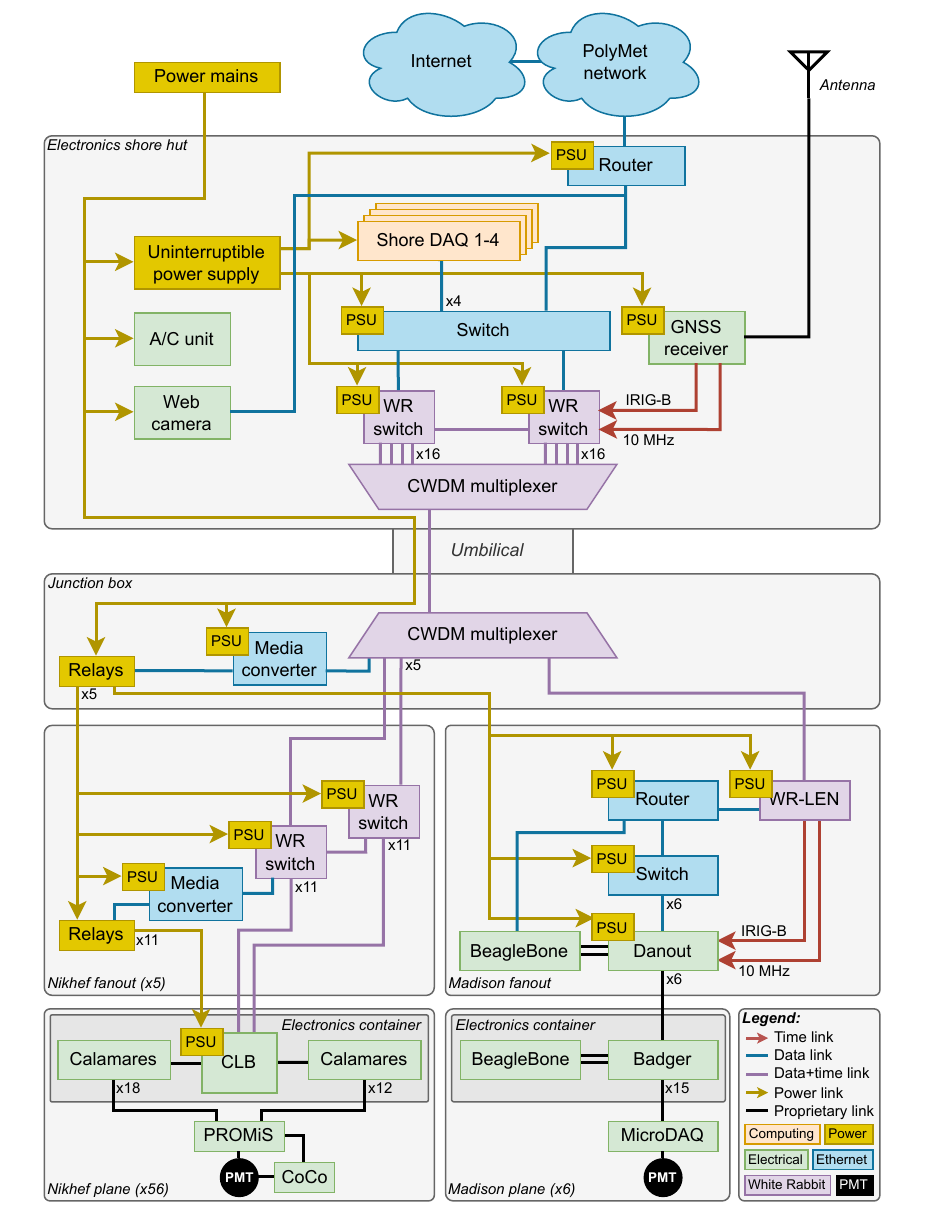}
    \caption{Diagram of \daq~hardware, showing organization of instrumentation inside~\chipsfive and on the shore. While measured data generally flow from the bottom up, commands, timing signals and power flows down from the top.}
    \label{fig:daq:shore:diagram}
\end{figure}

Besides computers and network infrastructure, the electronics hut also housed a cooling unit and power distribution systems, which supplied mains voltage to both shore huts as well as the deployed detector. While majority of devices were connected to an unprotected power line, which was potentially susceptible to outages, computers and key network elements drew power through an uninterruptible power supply~(UPS). In the event of a power outage this provided at least~\qty[round-precision=0]{15}{\minute} of battery capacity, which was sufficient for a graceful shutdown of any ongoing \daq~operations. A comprehensive overview of \daq~hardware used by~\chipsfive is shown in \Cref{fig:daq:shore:diagram}.

\section{Software} 
\label{sec:daq:software} 

After traversing the topology of \daq~hardware described in the previous section, measurements from~PMTs converged at 4~computers situated inside the electronics hut. These computers, labelled \textit{Shore~\daq~1-4}, hosted purpose-built software that constituted the next stage of the \daq~system. While it was desirable to deploy majority of programs to these machines given their relatively close proximity to the detector, it is worth noting that various components relevant to~\daq also operated in other locations. Firstly, CLBs~and~MicroDAQs inside the detector were programmed with dedicated firmware, which is beyond the scope of this work~\cite{aiello2019km3net,bendfelt2019microdaq}. Secondly, Fermilab provided a persistent storage facility for acquired data, which was planned to be used for physics analysis in later stages of the project. A \daq~subsystem was responsible for periodic data transfers from shore machines to this database. And finally, Fermilab infrastructure also hosted parts of \chipsfive~Timing Distribution System (TDS), which is discussed in other literature~\cite{manek2022low}.

Following conventional \daq~paradigm, the shore computers were programmed to perform three basic tasks:%
\begin{enumerate}
    \item
    Monitor the state of the detector.
    
    \item
    Control auxiliary electronics and detector instrumentation.

    \item
    Receive, process and store data from~PMTs.
\end{enumerate}

These tasks were implemented by a suite of single-purpose C++~programs~\cite{chips2019daq}, which were designed to function together in concert, much like detector hardware. This architecture choice had several advantages. Since each program had a clearly defined role and purpose, source code was relatively simple to write, and easy to inspect and verify. Furthermore, compartmentalization allowed \daq~programs to be individually instrumented for monitoring and independently reset upon crashing. Fault tolerance was further promoted by running software on only~2 out of~4 shore computers at any one time, leaving the other pair available as backups in case of failure or service interventions. To protect systems from hard drive faults, all persistent local storage was consolidated in a redundant array of inexpensive disks~(RAID)~\cite{patterson1988case}.

Shore computers were provisioned with Community Enterprise Operating System~(CentOS) Linux~7~\cite{centos2014}, which was the successor of Scientific Linux~6. To minimize system overhead, \daq~programs were deployed on `bare metal' without virtualization and executed continuously as daemons~(background processes). This allowed software to be supervised within the \texttt{systemd} service manager~\cite{systemd2010}, which was responsible for state tracking and automatic recovery from failures or degenerate states.

\subsection{Monitoring} 
\label{sec:daq:monitoring} 

To correctly control the detector following its deployment, it was important for \daq~operators to have complete and accurate representation of its state in real-time. This motivated development of a \textit{monitoring system}, which collected, visualized and archived various key indicators of interest over time. In addition, this system also dispatched urgent alerts whenever observed quantities exceeded specified tolerances or anomalies occurred with respect to established baseline.

The monitoring system used in~\chipsfive was based on Elasticsearch~\cite{elasticsearch}, which is an open-source document-based database system with broad applications in the field of infrastructure monitoring. This product was ideally suited for the task since its distributed architecture efficiently enabled periodic and simultaneous collection of structured time-ordered data from a wide range of sources. Based on desired use cases, Elasticsearch permits users to define indices that determine internal data organization strategy for fast retrieval, which is facilitated through Elasticsearch Query Language~(ESQL). In addition, Elasticsearch is accompanied by a diverse set of tools and integrations, collectively known as the Elastic Stack (or ELK~Stack)~\cite{elasticstack}, which readily implement compatibility with frequently used data sources and visualization methods.

For~\chipsfive, the Elasticsearch database engine was hosted from a single shore computer. Even though its underlying technology supported distributed operation over multiple machines, expected data throughput was sufficiently low to justify this simpler, more compact installation. Under normal operation the software exposed an application programming interface~(API) on the \daq~network, which received new information from data sources throughout the detector, as well as visualization requests and ESQL~queries from operators.

Data committed to the monitoring system came from several groups of sources. Firstly, the most basic information came from periodic health checks, which were directed from a Heartbeat process to all devices connected to the \daq~network. A single such check probed roundtrip latency of a conventional Internet Control Message Protocol~(ICMP) Echo~Request/Reply exchange. In addition, all Linux~computers (shore machines and BeagleBones) ran additional Metricbeat processes, which collected system logs, recorded state of all managed services (including \daq~programs) and sampled hardware counters. Similarly, compatible network elements (routers, switches and WR~switches) were configured to export their state to Elasticsearch over Simple Network Management Protocol~(SNMP).

Secondly, all \chips \daq~programs were instrumented to collect and publish their performance indicators through Elasticsearch~API. This information included detailed condition of all detector planes, local meteorological data (temperature, atmospheric pressure and relative humidity), the state of the GNSS~time source, number of hits registered from individual~PMTs, TDS~metrics, and \numi~operations data scraped from Fermilab Accelerator Division~(AD) website.

Finally, Elasticsearch consumed hit monitoring messages generated by software running on-board~CLBs and BeagleBones. These packets served as an independent integrity check throughout runs, and therefore needed to be archived for quality control purposes. Their significance is fully explained in \Cref{sec:daq:ingest}.

\begin{figure}
    \includegraphics[width=\columnwidth]{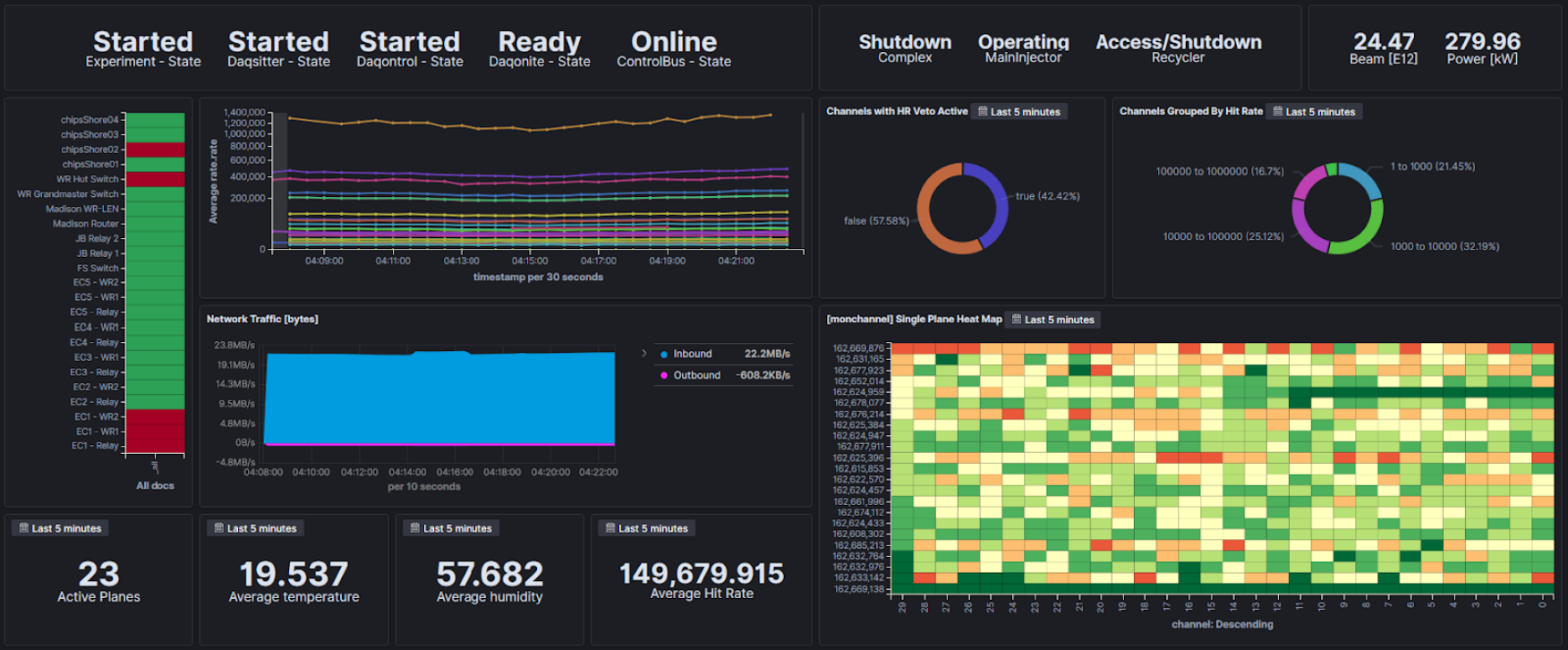}
    \caption{Monitoring dashboard visualized by Kibana, showing PMT~hit rates as well as state of the FSM and the \numi~beam reported from Fermilab. Plots and gauges are fed by data retrieved from Elasticsearch through ESQL~queries. Screenshot courtesy of J.~Tingey.}
    \label{fig:daq:monitoring:dashboard}
\end{figure}

Information accumulated by the \chipsfive monitoring system was presented to operators though the Kibana frontend~\cite{kibana}. In addition to issuing ESQL~queries and visualizing their outputs, this versatile web application was programmed with a collection of dashboards that topically compiled plots relevant to frequently performed \daq~operations. One such dashboard is shown in \Cref{fig:daq:monitoring:dashboard}. Kibana also included rule-based alerting mechanism, which was directed at Slack~channels and email addresses of \daq~operators.

\subsection{Detector control} 
\label{sec:daq:control} 

With accurate information supplied by the monitoring system, \daq~operators were able to inspect the state of individual components of the detector, and determine future actions. Expeditious and reliable implementation of their decisions was the purpose of the \textit{detector control system}.

Since the number of controllable elements in the first phase of \chipsfive~deployment was on the order of hundreds with planned future upgrades reaching thousands, it was viewed as intractable to control each component individually. This motivated division of the control system into three subsystems: a hardware control system~(HCS) that interacted with detector hardware, a finite state machine~(FSM) that tracked dependencies and orchestrated equipment, and a control bus that facilitated communications. \daq~operators typically issued high-level commands to~FSM, which in turn generated more concrete instructions to~HCS that carried them out. Messages exchanged between operators,~FSM and~HCS were dispatched through the control bus.

Representing the lower stage of the control system, HCS~directly interacted with a diverse set of components, such as~CLBs, BeagleBones and electrical relays. To this end, it implemented device-specific control protocols that exposed~APIs of parametrized actions and acknowledgements. Included among these were probing operations, which were regularly executed during health checks and forwarded to Elasticsearch, providing a source of monitoring information. Control exchanges with hardware were transmitted through \daq~network over Transmission Control Protocol~(TCP). To prevent network congestion from disrupting these communications, fixed minimal bandwidth for control data frames was allocated in network infrastructure. In addition, controlled equipment was isolated in multiple virtual local area networks~(VLANs).

\begin{figure}
    \includegraphics[width=\columnwidth]{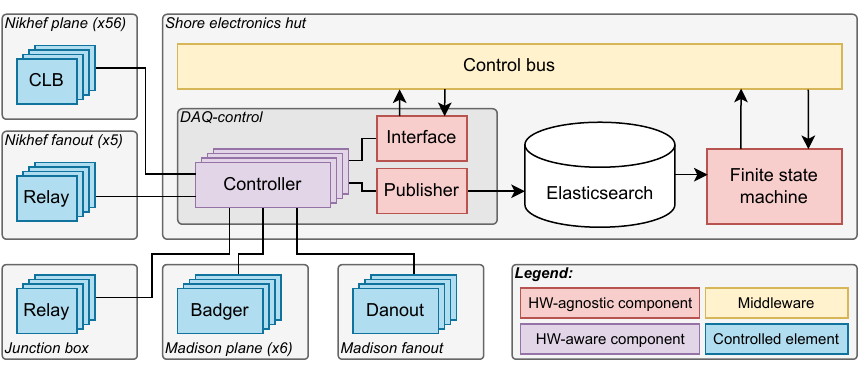}
    \caption{Diagram showing interaction of monitoring and control systems with controlled elements of~\chipsfive. Separation between hardware-aware~(violet) and hardware-agnostic~(red) components is indicated by color.}
    \label{fig:daq:control:diagram}
\end{figure}

It is worth noting that due to distinctiveness of instrumentation carried by Nikhef and Madison planes, their components responded to different low-level instructions that often performed similar actions. To prevent unnecessarily increased complexity in higher stages of the control system, HCS~acted as a hardware abstraction layer and presented a unified control interface for all planes towards FSM. This is illustrated in \Cref{fig:daq:control:diagram}. Implementation of control protocols was consolidated in a \textit{DAQ-control}~service, which was hosted at one of the shore computers; by convention the same instance that handled monitoring.

Higher up in the control system, FSM~implemented finite-state automata that tracked the \chipsfive~detector and all of its components. This method was selected because of its formal verifiability through model checking, high reliability and wide adoption in commercial industrial applications, which typically employ supervisory control and data acquisition~(SCADA) systems for the same task. FSM for~\chipsfive was built around the open-source TinyFSM~package~\cite{tinyfsm2012git}. It closely modelled detector hardware as a tree graph, where nodes represented individual components and directed edges connected nodes with a dependency relationship. In this graph, each node was characterized by a discrete state that described its configuration, health and readiness for \daq~operations. The set of all possible states was identical across all nodes, which allowed states to propagate upward along the topology of the graph according to a simple rule: if all children of a node shared the same state, the node attained that state. Conversely, if a node had children in multiple different states, it was marked as inconsistent, which usually attracted operator attention. The state of the entire detector was understood to be the state of the root of the tree.

To accurately determine the state of each node, FSM~continuously consumed information from the monitoring system. Based on that, \daq~operators were presented with a range of available actions that would initiate transitions to other reachable states. Each such action was applicable to individual nodes as well as larger subtrees of the graph. Once a state transition was requested, FSM~generated messages to all affected nodes through the control bus, which incorporated low-level instructions to~HCS that typically rendered nodes inconsistent for a short period of time. Once HCS~registered command acknowledgments, information retrieved through the monitoring system permitted~FSM to complete the state transition, eventually reaching a consistent state. A similar process occurred when failures were detected, with the notable distinction of the target state.

\begin{figure}
    \centering
    \includegraphics[width=0.7\columnwidth]{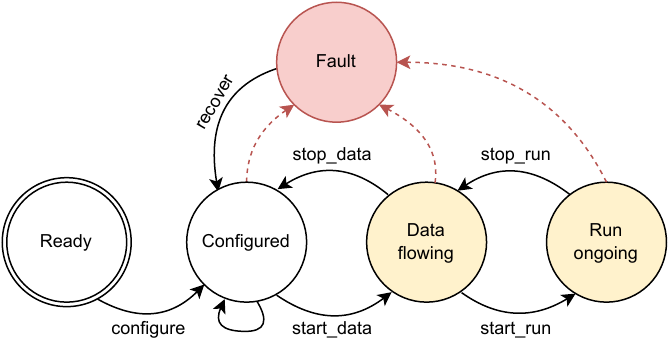}
    \caption{Simplified state diagram of the \chipsfive~FSM. The initial state is marked with a double border, whereas states where measurements flow through \daq~infrastructure are highlighted in yellow. Solid black and dashed red arrows indicate successful and failed state transitions, respectively.}
    \label{fig:daq:control:fsm}
\end{figure}

\Cref{fig:daq:control:fsm} shows a diagram of FSM~states relevant to \daq~operations. From the initial state, detector components underwent a batch configuration procedure, which applied calibrated cable delays, high voltages and digitization thresholds to individual~PMTs. In addition, it programmed measurement parameters and directed data flow at desired sinks. Successful configuration transitioned the detector to the `configured' state, primed to begin measurement. To begin registering PMT~hits, the operator moved to the `data flowing' state, which started ingest workers on the receiving side and opened shutter in~CLBs and BeagleBones. At that point, the detector became sensitive and \daq~network started experiencing increased traffic. This sometimes included short-lived surges (empirically, up to~\qty[round-precision=2]{5.66}{\giga\bit/s}), particularly when detector volume had been recently exposed to light. After a brief delay, which allowed data rates to stabilize, the operator commanded~FSM to transition to the `run' state that signalled ingest workers to enable data processing and collection. When this state was reached, measurements started accumulating as data files in persistent storage. To return the detector back to insensitive state, where its configuration could be adjusted, operators followed a reversed sequence of state transitions.

The communication channel that facilitated command input to FSM and connected FSM to HCS, was a brokerless control bus based on the NNG~middleware~\cite{nng2016git}. This open-source package was selected because of its minimalist, lightweight implementation and focus on increased robustness in production environments. Datagrams dispatched through the control bus were transmitted as reliable broadcasts through UNIX~domain sockets, where all bus participants acknowledged message reception.

\subsection{Data ingest} 
\label{sec:daq:ingest} 

When data flowed through \daq~network during runs, streams of information generated by planes were directed at multiple sinks, collectively known as \textit{data ingest}. Unlike monitoring and control systems, which aimed to provide low latency and high reliability, data ingest focused on efficient handling of large data volumes with a degree of tolerance for data loss.

It is worth noting that high throughput was a direct consequence of the design decision to use self-triggering in~\promis and~\microdaq. At a later point in the project, TDS~delivered the capability to trigger~\microdaq{}s externally by accelerator spills, which in part alleviated this effect. Nevertheless, this upgrade achieved only a modest reduction of overall data rates due to disproportionately small fraction of Madison planes in the detector. For that reason, efficiency under high data rates remained a priority even with external triggering.

At the level of~CLBs and BeagleBones, PMT~hits were aggregated into packets called \textit{optical data}. Representing the primary data output of a single plane, these messages were at most \qty[round-precision=1]{8.7890625}{\kilo\byte}~large\footnote{This was defined by setting Ethernet maximum transmission unit~(MTU) to~\qty[round-precision=0]{9000}{\byte}, in contrast to conventional~\qty[round-precision=0]{1500}{\byte}, for so-called jumbo data frames.}, and carried compressed~ToA, ToT~and channel address for every hit. Optical data were transmitted over User Datagram Protocol~(UDP), which is susceptible to data loss under certain conditions. To detect and track this undesirable effect throughout runs, CLBs~and BeagleBones generated a secondary data output, known as \textit{monitoring data} (not to be confused with data gathered by the \chipsfive~monitoring system). Monitoring packets were emitted every~\qty[round-precision=0]{10}{\milli\second} and contained metadata accompanied by snapshots of hit counters, as seen by the sender. By keeping independent hit counters and regularly comparing their values with these snapshots, the receiver was able to detect and quantify the fraction of dropped optical datagrams.

Under normal operation, packets from planes converged at~2 out of~4 shore machines. To fully utilize available processing capacity, a single machine was dedicated to each type of data, leaving the other pair as backups. Monitoring packets were received by the \textit{DAQ-monitor}~service, which decoded and forwarded hit counts to the monitoring system. This provided operators as well as hit handling software with a near real-time integrity check. In concert, optical data was directed at the \textit{DAQ-onite}\footnote{The name `DAQ-onite' was inspired by taconite, which is a type of iron ore mined in the region surrounding the \chipsfive~detector site.} service, which was responsible for producing measurement files compatible with the ROOT~analysis framework~\cite{brun1997root}.

Since both ingest services were designed to process data rates on the order of~\qty[round-precision=0]{10}{\giga\bit/\second}, their C++~implementation utilized multi-threading and asynchronous input/output~(ASIO) operations, which aim to avoid program suspension (known as \textit{blocking}) during time-consuming data access. For this reason, both programs interacted with network sockets and files through Boost.Asio~\cite{boost2020asio}. This open-source package was selected because of its focus on high throughput, well-characterized performance and abundance of adjustable parameters that permitted tunable scaling given specific computing hardware.

\begin{figure}
    \includegraphics[width=\columnwidth]{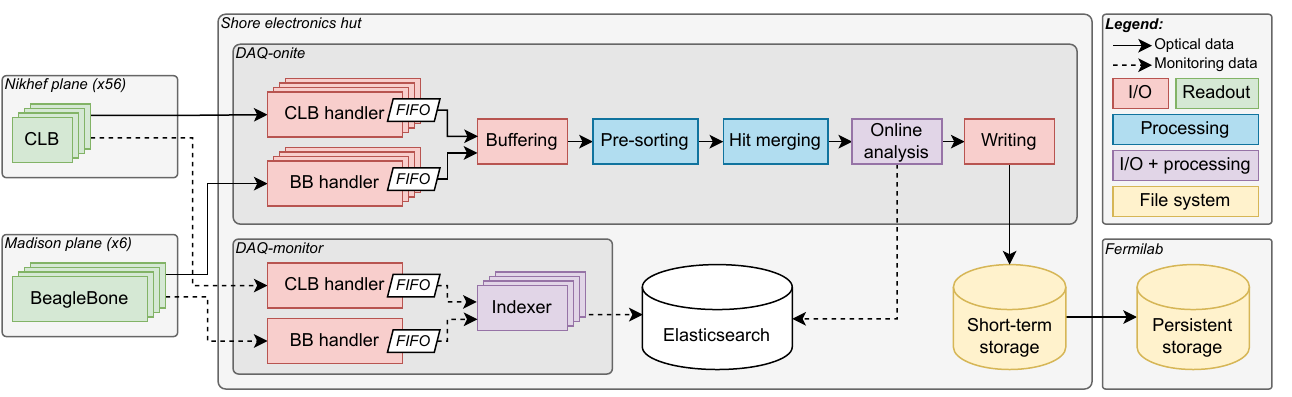}
    \caption{Diagram showing flow of optical and monitoring data from planes to ingest services and storage facilities. Here `FIFO' abbreviates \textit{first in, first out}, which is a lockless data structure utilized to relay data between processing threads.}
    \label{fig:daq:ingest:diagram}
\end{figure}

The internal architecture of an ingest service was divided into a hardware-oriented frontend, which consisted of packet handlers for Nikhef and Madison planes, and application-specific backend. This is illustrated in \Cref{fig:daq:ingest:diagram}. To minimize overhead, multiple parallel instances of frontend handlers were hosted in a thread pool that constituted a Boost I/O~service. Each handler consumed UDP~datagrams, decoded them and pushed their contents into a lockless memory data structure, which was continuously evacuated by the backend on another thread. In the case of DAQ-onite, the backend was a data handler that consolidated PMT~hits from planes, performed online analysis and saved outputs to ROOT~files. This was later upgraded to associate hit~ToA with incoming spill triggers, delivered by TDS from Fermilab accelerator complex~\cite{manek2022low}. In DAQ-monitor, the backend was an indexer that adapted and transmitted hit counts to Elasticsearch as structured documents.

\subsection{Hit merging and event building} 
\label{sec:daq:online_analysis} 

To complement immediate feedback from monitoring packets, DAQ-onite provided operators with data quality indicators through the monitoring system. Since these quantities often probed properties of individual events (coincident groups of hits), the program needed to ensure that hits arriving in parallel were first correctly interleaved into a single sequence, which was monotonically ordered by hit~ToA. This was the purpose of the \textit{hit merging} stage that operated in real-time during runs.

\begin{figure}
    \centering
    \includegraphics[width=0.5\columnwidth]{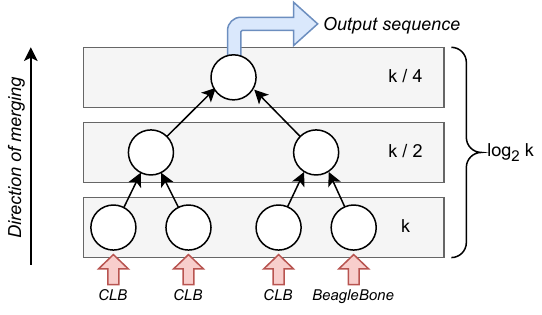}
    \caption{Iterative merging of $k$~hit sequences. Input sequences (red) are consolidated from the bottom of the binary tree upward, along the direction indicated by arrows. The output sequence (blue) is obtained from the root of the tree.}
    \label{fig:daq:online_analysis:hit_merging}
\end{figure}

The algorithm used to interleave hits was based on a conventional $k$-way merge strategy~\cite{thomascormen2009}, which was previously used with success in the \km3net~project. This method modelled $k$~input hit sequences as leaves of a binary tree. The algorithm independently processed leaf pairs in passes that linearly enumerated both corresponding sequences and took hits from one or the other based on the lowest available~ToA. In result, each leaf pair in the tree gave rise to a new parent node, which corresponded to a sequence containing hits from both children. This way, the merging process repeated multiple times, halving the number of nodes at the topmost level in every iteration. After at most $\mathcal{O}(\log_2 k)$~steps, all input sequences were merged into a single root node that contained output of the algorithm. This is illustrated in \Cref{fig:daq:online_analysis:hit_merging}.

Even though the $k$-way merge algorithm was formally correct, it made several ambitious assumptions that needed to be examined before its practical application in~\chipsfive. For instance, the algorithm expected input hit sequences to be finite arrays of known size, whereas in reality hits were registered continuously over time. Given the scale of the detector and hardware constraints of \daq~computers, it was intractable to retain optical data in memory until the end of a run, and only merge afterwards. For that reason, inputs were buffered in batches of fixed size. Depending on hit rates, DAQ-onite typically filled a single batch within seconds. This allowed the merging algorithm to always work with fixed-size inputs. In later stages of the project, hit merging was upgraded to process batches simultaneously on multiple threads in parallel to further increase performance.

\begin{figure}
    \centering
    \begin{subfigure}{0.45\columnwidth}
        \includegraphics[width=\textwidth]{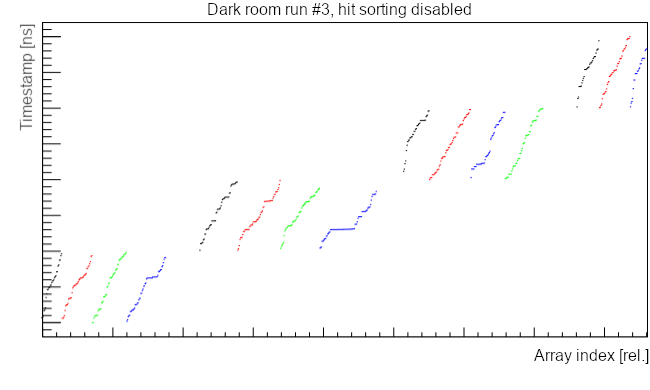}
        \caption{Received optical data from a CLB.}
        \label{fig:daq:online_analysis:unmerged}
    \end{subfigure}%
    \quad%
    \begin{subfigure}{0.45\columnwidth}
        \centering
        \includegraphics[width=\textwidth]{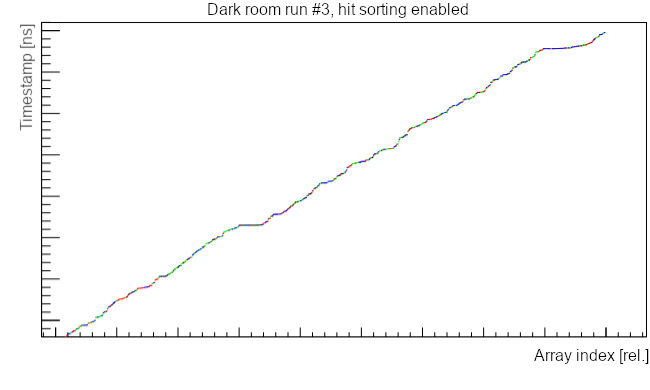}
        \caption{Pre-sorted batch, ready for hit merging.}
        \label{fig:daq:online_analysis:merged}
    \end{subfigure}
    \caption{Localized disorder found in CLB~optical data during a dark room measurement. Plots show hit~ToA as a function of hit number before~(a) and after~(b) the pre-sorting stage. PMT~channels are coded by point color.}
    \label{fig:daq:online_analysis:unmerged_merged}
\end{figure}

Another assumption made by the algorithm was that input hit sequences were independently well-ordered before they entered the merging stage. If this assumption held, a pair of sequences could be merged efficiently by a single linear pass. Sadly, even though CLB~optical data was mostly ordered overall, hit stream analysis revealed localized regions of disorder shown in \Cref{fig:daq:online_analysis:unmerged}. Since these regions aligned with channel boundaries, it is possible to speculate that CLB~firmware internally switched between channels in a round-robin scheme. While hits were correctly ordered within the scope of each channel, no effort was made to interleave hits between channels.

To address this hit disorder in DAQ-onite, additional pre-sorting stage was introduced between buffering and hit merging. In the course of pre-sorting, all $k$~input hit sequences were independently processed with insertion sort~\cite{donaldknuth1998}. Even though this algorithm is considered suboptimal for general use cases due to worst-case complexity of~$\mathcal{O}(n^2)$ given $n$~hits, its application here was well-motivated. Since complexity of insertion sort is proportional to the degree of disorder of its input, applying it to a nearly ordered hit sequence typically consumed only negligible fraction of computing resources. In an ideal case, where the input was already perfectly ordered, complexity of the algorithm was reduced to appreciable~$\mathcal{O}(n)$. In addition, the use of insertion sort in pre-sorting also brought practical benefits for Madison planes. Even though Field~Hub~App internally interleaved hits from \microdaq{}s before transmission, high hit rates sometimes tested limits of BeagleBone processing capacity. In such situations, the existence of pre-sorting allowed compute loads to be alleviated by seamlessly transferring them upstream to DAQ-onite without jeopardizing correctness of collected data.

Moving past the merging stage, hits were clustered into events by comparing differences of successive~ToA, and grouping hits that fell into an adjustable time window on the order of nanoseconds. This was also the point where events were classified as cosmic if they contained any hits from veto~PMTs, and where data quality indicators were calculated. Following that, data were committed to persistent storage in the ROOT~file format.

\subsection{Offline analysis} 
\label{sec:daq:offline_analysis} 

Produced measurement files were temporarily retained at shore computers. Due to their relatively low size, locally available storage capacity sufficed for several weeks of continuous acquisition, if necessary. Under normal operation, however, data were regularly transferred over the~Internet to a permanent storage facility at Fermilab. From there, archived measurements were subject to \textit{offline analysis}, which was planned to be implemented in batch computing jobs submitted to Open~Science~Grid~\cite{pordes2008open}. Consistently with design principles of the \chips~project, offline analysis explored novel approaches to neutrino event simulation, reconstruction and classification, which had potential to reduce complexity and computing loads while maintaining or possibly improving accuracy.

\begin{figure}
    \centering
    \begin{subfigure}{\columnwidth}
        \centering
        \includegraphics[width=0.5\linewidth]{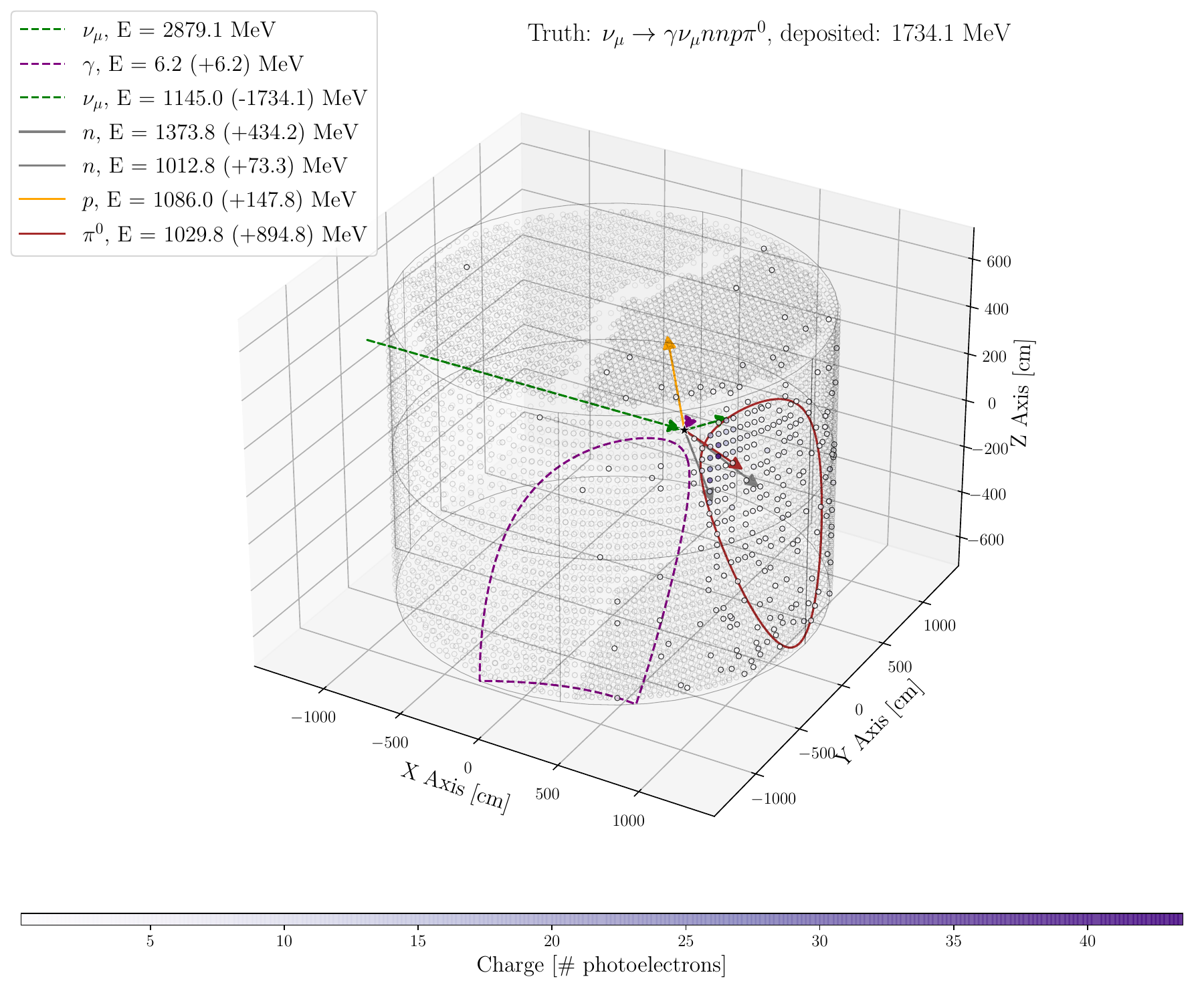}%
        \includegraphics[width=0.5\linewidth]{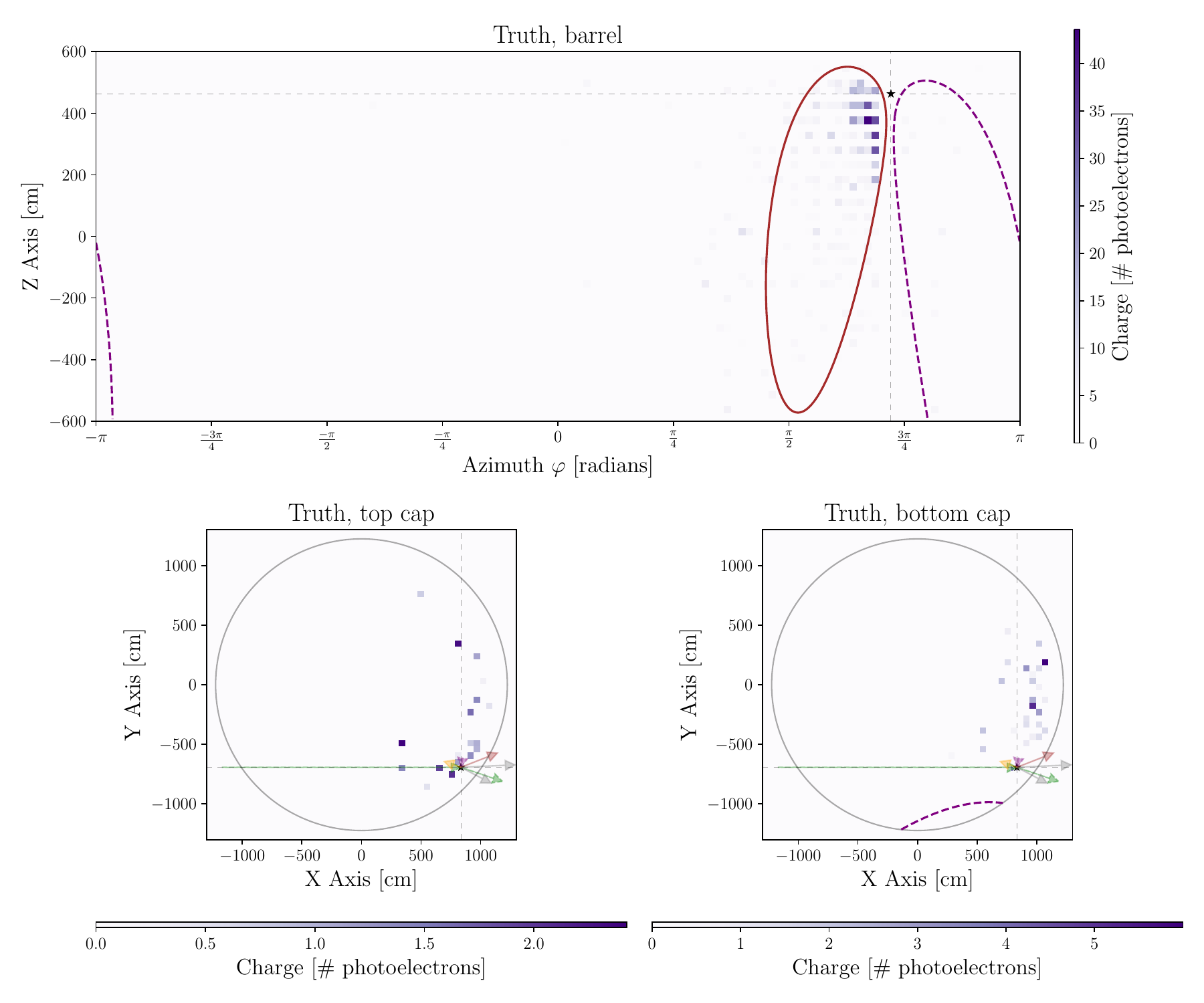}
        \caption{Visualization of the simulated ground truth.}
        \label{fig:daq:offline_analysis:track_reconstruction:truth}
    \end{subfigure}

    \bigskip
    \bigskip

    \begin{subfigure}{\columnwidth}
        \centering
        \includegraphics[width=0.5\linewidth]{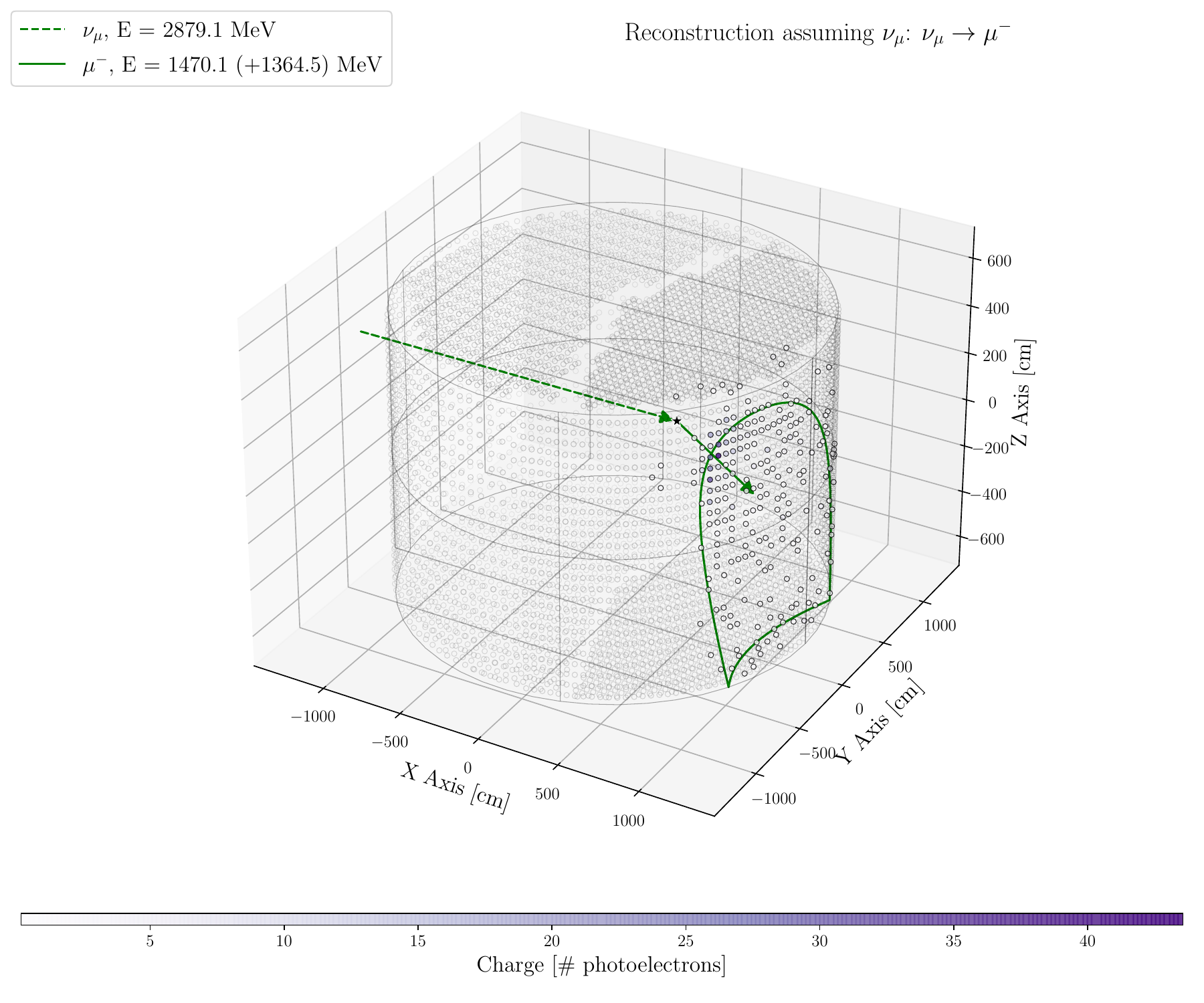}%
        \includegraphics[width=0.5\linewidth]{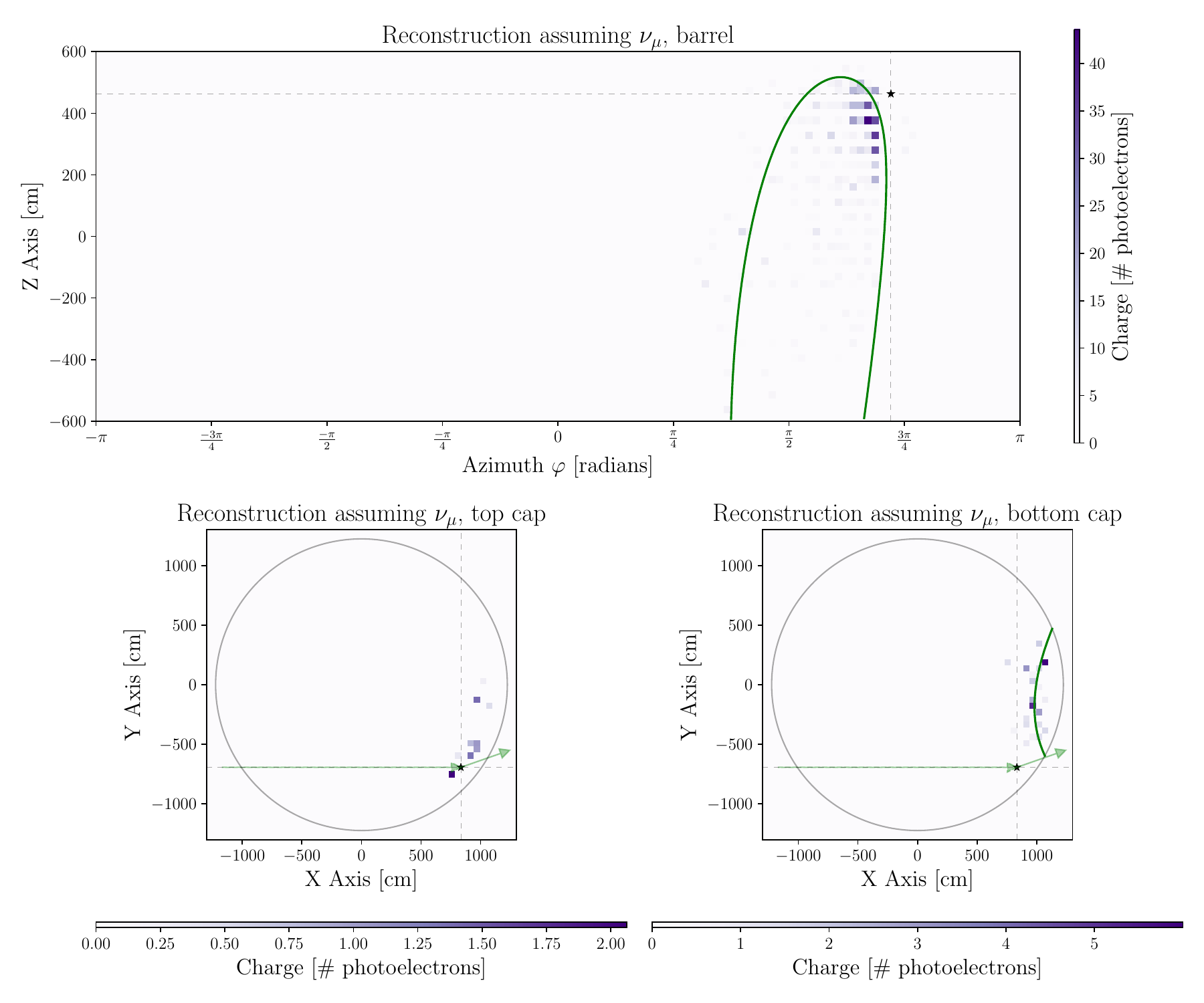}
        \caption{Most probable hypothesis found by a likelihood-based fit.}
        \label{fig:daq:offline_analysis:track_reconstruction:reco}
    \end{subfigure}

    \bigskip

    \caption{Example of likelihood-based fitting. Plots display a simulated \Pnum~interaction~(a) and its reconstruction~(b), visualized as charge collected by PMTs~(color axis) and Cherenkov cone sections~(contours). While the left view shows a perspective of the detector geometry, two-dimensional maps on the right correspond to inner faces of the detector. Particle trajectories are indicated by arrows, the interaction vertex is marked by a star.}
    \label{fig:daq:offline_analysis:track_reconstruction}
\end{figure}

A comprehensive description of a likelihood-based event reconstruction algorithm is available in reference~\cite{blake2016chips}. This method required prior generation of a sizeable data sample using the WCSim~package~\cite{wcsim2020}, which was adapted for \chips~geometry from its original application in \t2k and~LBNE. In its implementation~\cite{chips2019sim,chips2019reco}, beam and cosmic events, supplied by the~GENIE~\cite{andreopoulos2010genie} and CRY~\cite{hagmann2007monte} generators respectively, produced digitized PMT~response that was simulated using the GEANT4 toolkit~\cite{agostinelli2003geant4}. When presented with a real-world event, the reconstruction algorithm began by identifying Cherenkov~rings in a geometric hit map using circular Hough~transform~\cite{mukhopadhyay2015survey}. In subsequent iterations, seeded tracks were fitted by maximizing likelihood of observed hits given previously simulated detector behaviour. A result of such a fit is shown in \Cref{fig:daq:offline_analysis:track_reconstruction}. The final stage of the algorithm employed two feed-forward artificial neural networks~(ANNs) for particle identification~(PID). At input, these networks accepted vectors of various manually selected features, such as hit count ratios yielded by the preceding stage. At output, the networks produced a classification that labelled events as electron-like, muon-like or NC~interactions.

While likelihood-driven optimization presented a viable method of event reconstruction, its implementation was inherently based on empirical observations and hand-engineered features that provided no guarantee of optimality. This motivated development of alternative methodology, which at its core employed convolutional neural networks~(CNNs). Since these models perform parametrized feature extraction in initial layers of their network architecture, their training was expected to overcome the optimality issue by favouring features best-suited to the task.

\begin{figure}
    \centering
    \includegraphics[width=\columnwidth]{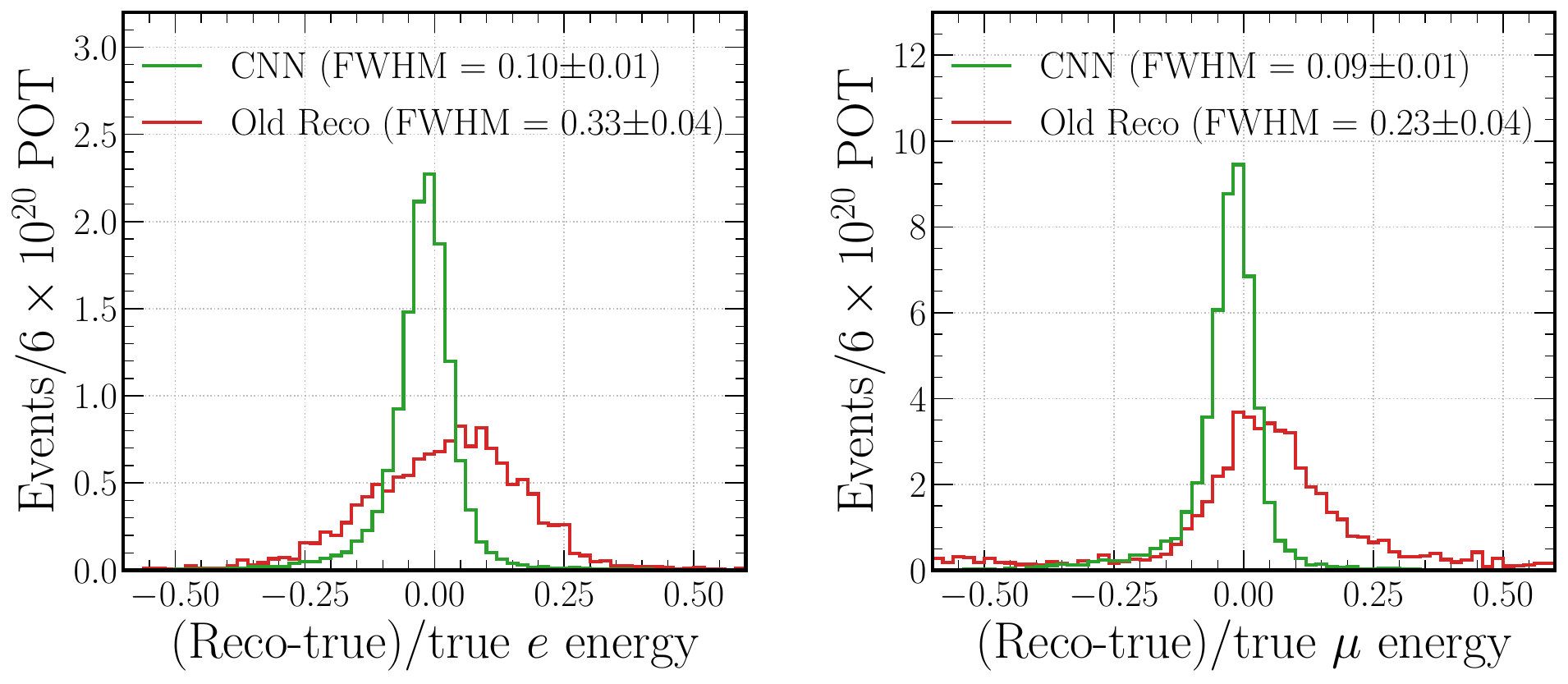}
    \caption{Comparison of neutrino energy reconstruction accuracy by likelihood-based fitting~(red) and a CNN~(green). Plots show the difference between reconstructed and true energies of quasi-elastic~CC \Pnue~(left) and \Pnum~(right) interactions on a sample of \num[round-precision=0]{400e3}~beam events. The result was taken from reference \cite{tingey2023neutrino}.}
    \label{fig:daq:offline_analysis:cnn}
\end{figure}

Event reconstruction techniques based on~CNNs are discussed in reference~\cite{tingey2023neutrino}. In many aspects, these methods built on their likelihood-based counterparts by adapting their simulation software for production of training datasets. Simulated events were arranged into geometric hit maps that directly encoded digitized~ToT and~ToA, complemented by an additional parameter map that was calculated by Hough transform. These two-dimensional maps were interpreted as images and fed into a deep network known as \textit{chipsnet}, which was largely based on the widely adopted VGG~architecture~\cite{simonyan2014very}. This gave rise to three categories of networks, distinguished by interpretation of their output layer: the \textit{cosmic rejection network} that classified event origin as beam or cosmic, the \textit{beam classification network} that further tagged beam events as electron-like, muon-like or NC~interactions, and the \textit{neutrino energy network} that performed regression of neutrino energies. In evaluation, these models were shown to yield appreciable improvements over likelihood-based fitting. This can be for example demonstrated in terms of errors in neutrino energy reconstruction, which are compared in \Cref{fig:daq:offline_analysis:cnn}.

\section{Conclusion} 
\label{sec:conclusion} 

The goal of the \chips project was to show that affordable neutrino detector technology could be implemented using commercially available components, rather than their more expensive purpose-built counterparts. The explored domain of cost reduction strategies included novel design concepts, construction methods and \daq~technology. To this end, modern instrumentation was developed for use in the recent \chipsfive detector prototype, which was based on two distinctive but functionally equivalent sets of hardware. Inspired by the architecture of \chipsfive itself, this \daq~system adopted a modular approach that allows a basic detector to be composed from a small number of interconnected~POMs, which can be easily serviced and readily scaled up. In spite of its versatility, the presented technology does not make undesirable compromises on data quality. Specifically, its sub-nanosecond time resolution is sufficient to reconstruct directional sources of Cherenkov light within detector volume. Furthermore, its compatibility with standardized timing distribution systems permits synchronization and interoperability with other similar detectors.

Hardware components of the presented \daq~solution are accompanied by a suite of high-performance programs that are capable of monitoring, controlling and reading out instruments over a conventional Ethernet network. This software was designed with a robust and parallel architecture, which allows it to be efficiently deployed in distributed computing facilities in order to orchestrate detectors on the scale of \chipsfive. In addition to decoding and saving received data, the software is able to recognize coincident events in real-time and associate this information with triggers generated by neutrino spills. Offline event reconstruction is facilitated by two alternative algorithms: a conventional likelihood-based fit and a more modern method that delegates feature extraction as well as event classification to a~CNN. Both of these techniques have shown acceptable accuracy when evaluated with simulated events.

Even though the \chipsfive~prototype was ultimately unable to observe \numi~neutrinos due to deployment issues, its development revealed potential for appreciable cost reductions achieved by transition to standardized mass-produced components and modular architecture. Furthermore, during the course of the project the presented \daq~system successfully passed a number of commissioning tests, indicating that the instrumentation was functional and suitable for scientific use at the intended scale. The technology is currently reused in other ongoing experiments.

\section{Acknowledgements} 
\label{sec:acknowledgements} 

This work was supported by Fermilab; the Leverhulme Trust Research Project Grant; U.S. Department
of Energy; and the European Research Council funding for the CHROMIUM project. Fermilab is operated
by Fermi Research Alliance, LLC, under Contract No.~DE-AC02-07CH11359 with the U.S. DOE.

The \chips collaboration would like to thank the Wisconsin IceCube Particle Astrophysics Center, University of Wisconsin–Madison, University College London, University of Alberta, University of Minnesota Duluth and Aix-Marseille University. Finally, the \chips collaboration would like to thank PolyMet and Cleveland-Cliffs for hosting the experiment.
\begin{adjustwidth}{-\extralength}{0cm}

\reftitle{References}


\bibliography{article}


\PublishersNote{}
\end{adjustwidth}

\end{document}